\documentclass{emulateapj}
\slugcomment{accepted for publication in ApJ}

\usepackage{gensymb}
\usepackage{textcmds}
\usepackage{graphicx}
\shorttitle{Orbital dynamics of Exomoons in Planetary Close Encounters}
\shortauthors{Hong et al.}

\begin{document}

\title{Innocent bystanders: Orbital dynamics of exomoons during planet-planet scattering}
\author{Yu-Cian~Hong$^{1,2}$,Sean~N.~Raymond$^3$,Philip~D.~Nicholson$^1$, and Jonathan~I.~Lunine$^{1,2}$}
\email{ycsylva@gmail.com}
\affil{$^1$Astronomy Department, Space Sciences Building, Cornell University, Ithaca, NY 14853, USA\\
$^2$Carl Sagan Institute, Space Sciences Building, Cornell University, Ithaca, NY 14853, USA\\
$^3$Laboratoire d'astrophysique de Bordeaux, Univ. Bordeaux, CNRS, B18N, all{\'e}e Geoffroy Saint-Hilaire, 33615 Pessac, France}

\begin{abstract}

Planet-planet scattering is the leading mechanism to explain the broad eccentricity distribution of observed giant exoplanets. Here we study the orbital stability of primordial giant planet moons in this scenario. We use N-body simulations including realistic oblateness and evolving spin evolution for the giant planets.  We find that the vast majority ($\sim 80\mbox{—-}90\%$ across all our simulations) of orbital parameter space for moons is destabilized.  There is a strong radial dependence, as moons past $\sim 0.1 R_{Hill}$ are systematically removed.  Closer-in moons on Galilean-moon-like orbits ($<$ 0.04 $R_{Hill}$) have a good ($\sim 20\mbox{—-}40\%$) chance of survival.  Destabilized moons may undergo a collision with the star or a planet, be ejected from the system,  be captured by another planet, be ejected but still orbiting its free-floating host planet, or survive on heliocentric orbits as "planets".   The survival rate of moons increases with the host planet mass but is independent of the planet's final (post-scattering) orbits.  Based on our simulations we predict the existence of an abundant galactic population of free-floating (former) moons.

%\keywords{celestial mechanics $\mbox{--}$ exomoon $\mbox{--}$ orbital stability $\mbox{--}$ planet oblateness $\mbox{--}$ perturbation $\mbox{--}$ non-coplanar planetary systems}

\vspace*{3\baselineskip}

\end{abstract}

\newpage

\section{Introduction}
Given that each of the Solar system's giant planets hosts at least one large natural satellite, the presence of moons is anticipated around giant exoplanets.  The potentially diverse environments on exomoons, and the clues they may provide to planet formation models makes them subjects worthy of research.  

The potential habitability of exomoons is affected by a number of processes, including atmospheric dynamics, stellar illumination, tidal heating, planetary magnetic fields, orbital configurations, etc \citep{dobos,forgan,heller12,heller13a,heller13b,heller14a,hinkel,kaltenegger,tinney,williams}.  Exomoons have more paths to habitable configurations than planets.  Besides being in the liquid water habitable zone of the star \citep{kasting}, effects such as tidal heating provide an opportunity for moons outside the stellar habitable zone to be heated \citep{dobos,heller13a,reynolds,scharf}. 

Several techniques appear capable of detecting exomoons (see Section 6 of \citet{heller14c} for a review).  The transit technique has the prospect of observing sub-Earth sized moons \citep{agol,holman,kipping09a,kipping09c,kipping12,sartoretti,simon,kipping15,heller14b}, and has discovered a Neptune-sized exomoon candidate \citep{teachey}. Microlensing can detect moons down to 0.01 Earth masses \citep{bennett96,bennett02,han02,han08} and has discovered a sub-Earth mass exomoon candidate around a free-floating planet \citep{bennett14}.  Its broader sensitivity with heliocentric distances may yield a better chance of exomoon detection.  Direct imaging may be able to detect bright, tidally heated exomoons \citep{peters}.

The frequency of the occurence of exomoons depends on their formation and dynamical evolution.  The Hill stability criterion, Roche limit, stellar tidal stripping, tidal decay, planet migration, direct planet perturbation, and planetary close encounters are among the mechanisms that can destabilize the orbits of exomoons \citep{barnes,domingos,donnison,frouard,holman99,kane,payne,sasaki,namouni,spalding}.  The stability of distant satellites ($a_{s} > 0.1$ $R_H$ ), which would be classified as irregular satellites in the Solar System, in planet-planet scattering, are investigated by \citet{gong} and \citet{nesvorny}.  

In this paper we investigate the orbital dynamical behavior and the stability of primordial satellites as close-in as Io to Jupiter ($\sim$ 0.01 $R_H$), in scenarios where planets gravitationally scatter off each other.  Planet-planet scattering has been considered as the most viable candidate mechanism for explaining the prevalence of eccentric orbits of extra-solar planets and reproduced well their observed eccentricity distribution \citep{adams,chatterjee,ford,juric,lin,marzari,rasio,raymond,weidenschilling}.  Therefore, it is important to test the orbital stability of planetary satellites in this context.  In the planet-planet scattering scenario, planets are hypothesized to form closely packed, then they perturb each other and eventually enter an instability phase when they undergo orbit crossing and close encounters.  The instability is ended by the removal of some planets in the system by ejection or collision.  In the instability phase, the planetary satellites experience strong gravitational perturbations from the perturbing planets.  Various sources of perturbations also affect the satellites, such as secular perturbations from the perturbing planets (as opposed to the host planet of the satellites) , and stellar perturbations through the change of the host planet's spin and orbits in the instability phase.   The orbital parameter space of moons in this work covers close-in regions where planet oblateness plays a major role in moon stability, as an already well-established fact, and the absence of planet oblateness will cause unrealistic orbital instability effect \citep{hong15}. This work also simulates moons within and beyond the critical semi-major axis (0.04 Hill radii for Jupiter) of the planets where planet spin can affect moon stability \citep{tremaine09}, and moons up to 0.35 Hill radii where prograde moons can be stable.

Section 2 studies the inner working of planetary close encounters and discusses factors related to close encounters that affect moon stability,by simulating single planet mutual flyby events.  Section 3 discusses the orbital evolution of moons under various sources of perturbations, moon survival rate and its relevant factors, and the dynamical outcome of moons in planet-planet scattering in full simulations of length $1 \mbox{-} 100$ million years.

\section{Numerical method}
This work uses the N-body symplectic integrator \textit{Mercury} \citep{chambers}. All simulations use the Bulirsch-Stoer algorithm, the most accurate
 for simulating bodies that perturb each other closely, though the slowest in the package, in order to assure integration accuracy.  The conservative version of Bulirsch-Stoer integrator is used, which is appropriate in cases where the force does not depend on velocity.
 The spin and oblateness of the planet are adapted into the code in order to correctly simulate the stability of planetary satellites.

\subsection{Oblateness of planet}
The planets are simulated as oblate by an addition of the quadrupole moment ($J_2$) term to the gravitational force between point mass bodies.  This treatment ensures that the orbits of moons 
within the critical orbital distance from a planet are simulated correctly and that unexpected instability will not occur while the planet keeps the close-in moons' orbital angular momentum precessing rapidly \citep{hong15}.
Numerically this is done by adding a customized force term that accounts for the quadrupole moment contribution in the gravitational attraction of a planet to other planets and satellites in \textit{Mercury}.  Since the spin of the planet is allowed to change, the direction of the force from the planet's bulge follows the spin evolution of the planet.  However, the code assumes that the spin of gas giant planets can react to external torques instantaneously during a planetary close encounter, whereas in reality the timescale for a fluid body to react to torques might be longer than the close encounter time.

\subsection{Spin evolution of planet}
The spin of an oblate planet evolves under the influence of the central body and other planetary bodies, especially during a close encounter between planets.  In the planet-planet scattering scenario, the spin of a planet can
often evolve dramatically, in which the stability of its moons needs to be tested \citep{tremaine09}.
In this work, the spin evolution model for the planet uses an instantaneous equation of motion,
 in order to correctly simulate the effect of planetary close encounters on the spin of a planet.  
 The equation of motion for the spin evolution is derived in the following paragraphs.

 An extended mass receives different amounts of gravitational attraction from a distant point mass at  different radial distances within its interior, which results in a torque on the extended mass,
 causing its spin to evolve.  The force from a point mass M on an extended mass m is

\begin{equation}
       \vec{F} = \frac{-\,G\,M\,m}{r^2} \{\hat{r} - 3\,J_{2}\,(\frac{R}{r})^2\,[(5(\hat{r}\cdot\hat{s})^2 - 1)\,\hat{r} - 2\,(\hat{r}\cdot\hat{s})\,\hat{s}] + \cdots\},
\end{equation}

where G is the gravitational constant, r is the distance between the centers of m and M and $\hat{r}$ points from m to M, R is the radius of m,
and $\hat{s}$ the normalized spin vector of m \citep{hilton91}.
The torque per moment of inertia is 

\begin{equation}
        \frac{\vec{r}\times\vec{F}}{I} = \frac{d}{dt}\left(\Omega\,\hat{s}\right),
\end{equation}

, where $\Omega$ is the rotation period of the planet, which for simplicity is always assumed to be constant and equal to 10 hours, the same as for Jupiter.
The time evolution of the spin components can thus be described in the following equation

\begin{equation}\label{eq:dsdt_r}
       \frac{d\hat{s}}{dt}= \frac{-3\,G\,M\,m}{r^3} \frac{J_{2}}{\lambda\,\Omega} \left(\hat{r}\cdot\hat{s}\right)\left(\hat{r}\times\hat{s}\right),
\end{equation}

where $\lambda$ is the normalized moment of inertia. $\lambda$ is taken to be 0.25, close to that of Jupiter, in all simulations.  The above equation is derived assuming an instantaneous force and torque within which no variables are averaged over time.  It is therefore 
suitable for the planet-planet scattering case, where planets sometimes interact extremely closely $( ie. \sim0.001 AU)$, thereafter over a brief period of time  $(ie. \sim1yr)$.  As a side note, equations of spin evolution for secular problems such as the obliquity
evolution of Mars require the orbit to stay unchanged at least over the timescale of the body's orbital period:

\begin{equation}
       \frac{d\hat{s}}{dt}= \left(\frac{-3}{2}\frac{n^2}{\Omega}{J_2}{\lambda}\right)\left(\hat{l}\cdot\hat{s}\right)\left(\hat{l}\times\hat{s}\right)
\end{equation}

\citep{ward73,ward74,bills90,bills05}.  The above equation uses averaged quantities such as $\hat{l}$, which is the angular momentum averaged over an orbital period.  During a close encounter, in particular, such an assumption lacks accuracy because the small timescale of the variation of perturbation is sometimes comparable to the orbital period of the binary planet pair.  

Eq. (\ref{eq:dsdt_r}), with its dependence on the body's position relative to other bodies, is plugged into the Bulirsch-Stoer integrator alongside the integration for positions and velocities, and is updated once per time step.  The validity of the integrator is tested with the past obliquity evolution
of Mars.  The initial conditions are taken from \citet{quinn91} at epoch JD 2433280.5, with the obliquity of Mars $ = 25\degree$.  The simulation is 
integrated backwards and produces results with close resemblance to \citet{ward73}.

Unless otherwise specified, the initial settings of the simulations are configured as follows : 1) the integration error limit in orbital energy and angular momentum is $ 10^{-12}$, 2) the integration time-step is 0.1 days, in order to accurately integrate satellites with orbital periods as short as a few days like the Galilean satellites.

\section{Result: Single Close Encounter}\label{sec:single} 
We start by considering the effect of a single planet-planet encounter on the stability of one planet's moons.

\subsection{Simulation Setting}

  In each simulation of simulation set 1, the system contains a Sun-like star, a Jupiter-mass planet that hosts moons, and a perturbing planet.  The host planet is at 5 AU on a circular orbit.  Moons are massless test particles placed between 0.01 $\mbox{--}$ 0.35 Hill radii (thereafter $R_H$) from the host planet on circular and co-planar orbits.  Four moons are spaced out evenly in angular position at each of the 10 planetocentric semi-major axes.  The planets are set up to undergo close encounters (within each other's Hill sphere) by the following.  The perturbing planet is initially placed at 1.2 Hill radii from the host planet.  The parameters of the perturbing planet are the only variants across different simulations in set 1.  The perturbing planet's mass is equal to 0.1, 0.5, 1.0, or 2.0 $M_J$ (Jupiter mass).  Its starting velocity relative to the host planet ($V_{rel}$) ranges from 0.001 to 0.009 AU / day, and the direction of its velocity vector is restricted to lie within 60$\degree$ from the host planet's velocity vector.  The range of $V_{rel}$ is determined from all close encounters in a subset of planet-planet scattering simulations for $10^6$ yr.  The perturber's starting impact parameter (b) is 0.005 - 0.1 AU ( host planet radius = 0.00047 AU).  The impact parameter is set as a 3-dimensional sphere of vectors surrounding the center of the host planet.  All simulations last for 10 years.  Simulations in which planets collide with each other or in which the perturber starts with e $>$ 1 are excluded.  In most of the close encounters during the instability period, the latter configurations are rare.

\subsection{Results}

\begin{figure*}[!t]
    \centering
   \includegraphics[width=0.65\textwidth]{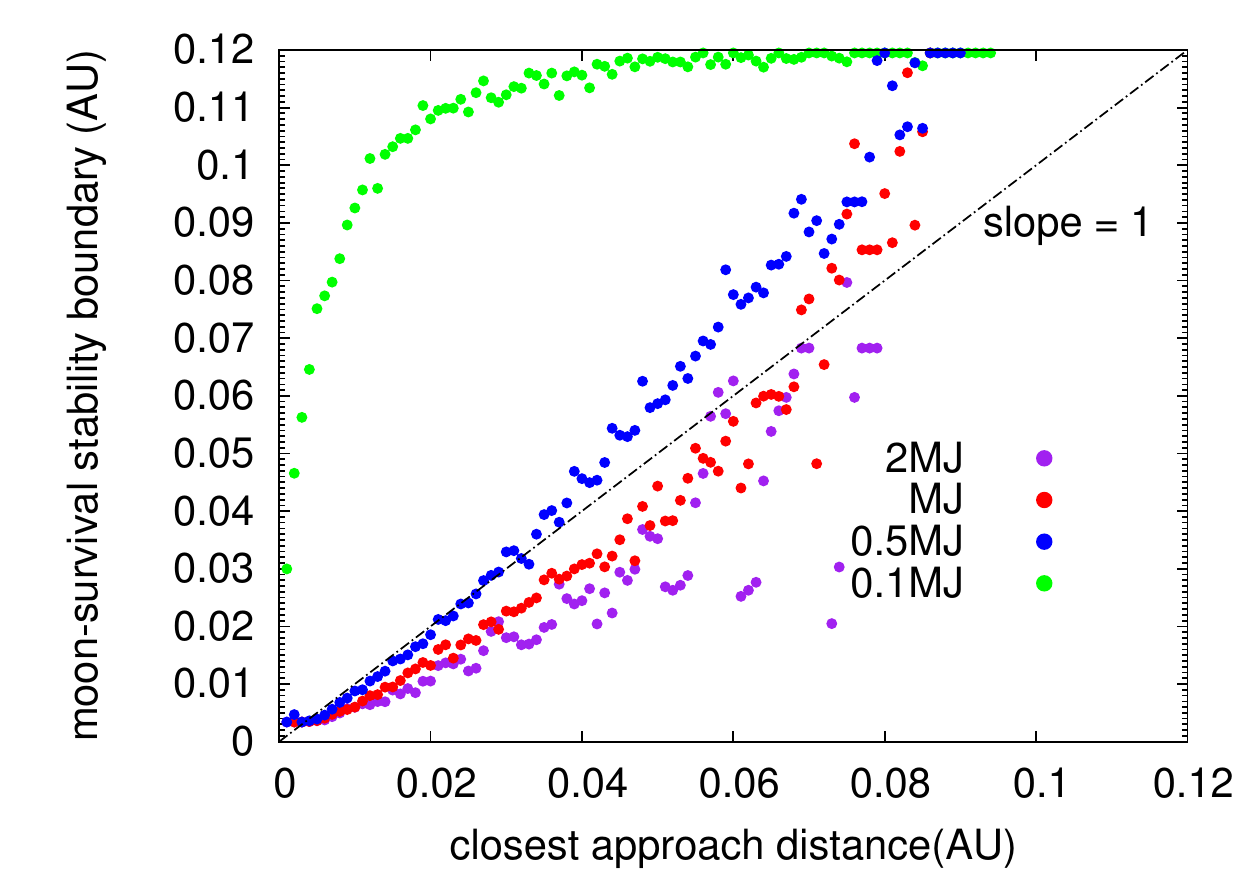}
  \caption{%
 Planet minimum close encounter distance vs. moon stability boundary.  The stability boundary is set where the most distant surviving moon with $e <$ 0.5 is located around the planet.  Dots with different colors represent simulations with different perturber mass as shown in the legend.
Each dot is obtained from taking averages of simulations with the same $d_{min}$.  The trend in the plots with perturber mass 0.5 $\mbox{--}$ 2 $M_J$ indicates that $d_{min}$ plays the most important role in determining the moon stability boundary, and the different slopes of each curve corresponding to different perturber masses shows that perturber mass is a secondary factor.  
  }\label{figk1}
\end{figure*}

The minimum close encounter distance $d_{min}$ plays a major role in determining moon stability.  From the initial conditions, it is mainly $\textit{b}$ that determines $d_{min}$ and thus moon stability.  $V_{rel}$ plays a minor role because the escape velocity of the host planet (0.035 AU / day) is $1\mbox{--}2$ orders of magnitude ($\sim 35\mbox{--}350$ times) greater than $V_{rel}$.  Figure \ref{figk1} shows the stability boundary of moons around host planets
that experience close encounters with different closest approach distances.  The stability boundary is set where at least 2 moons on the same planetocentric distance survive with e $\leqslant$ 0.5, because they
are anticipated to have high probabilities to survive a full simulation. As will be discussed in section \ref{sec:full}, 80$\%$ of the surviving moons have eccentricities under 0.5. The stability boundary
 for each $d_{min}$ is an average taken from all the host planets with the same $d_{min}$
and with a perturber having the same mass.  The larger $d_{min}$ is, the larger the stability boundary.  For simulations with perturber mass $0.5\mbox{--}2 M_J$, the stability limits in figure \ref{figk1} are positively correlated with $d_{min}$ in the curves, which demonstrates that $d_{min}$ plays the most important role.  The slope of the curves is determined by the perturber mass, demonstrating that the perturber mass plays the second most important role in determining the stability boundary of a planet.  The slope is roughly linear when the perturber mass is less than twice that of the host planet, but when the perturber mass reaches twice that of the host planet, the linear relation starts to fail.

Host planets with lower-mass perturbers have larger stability boundaries.  Simulations with a 0.5 $M_J$ perturber have stability boundaries close to or greater than $d_{min}$, and simulations with a 0.1 $M_J$ perturber have much larger stability boundaries than $d_{min}$.  For simulations with larger perturbers $(M_J$ and 2 $M_J)$, the stability boundary is $0.6\mbox{--}0.8$ times $d_{min}$ for $d_{min} < 0.05 AU$.  Except for simulations with a 2 $M_J$ perturber, there exists a boundary in $d_{min}$ beyond which almost all moons from $0.01\mbox{--}0.35 R_H (\sim 0.003\mbox{--}0.12 AU)$  are stable.  For simulations with a 0.1/0.5/1 $M_J$ perturber, when $d_{min}$ is greater than $\sim 0.05/0.1/0.1$ AU, the stability boundary lies close to 0.35 $R_H$.

\begin{figure*}[!t]
    \centering
   \includegraphics[width=0.65\textwidth]{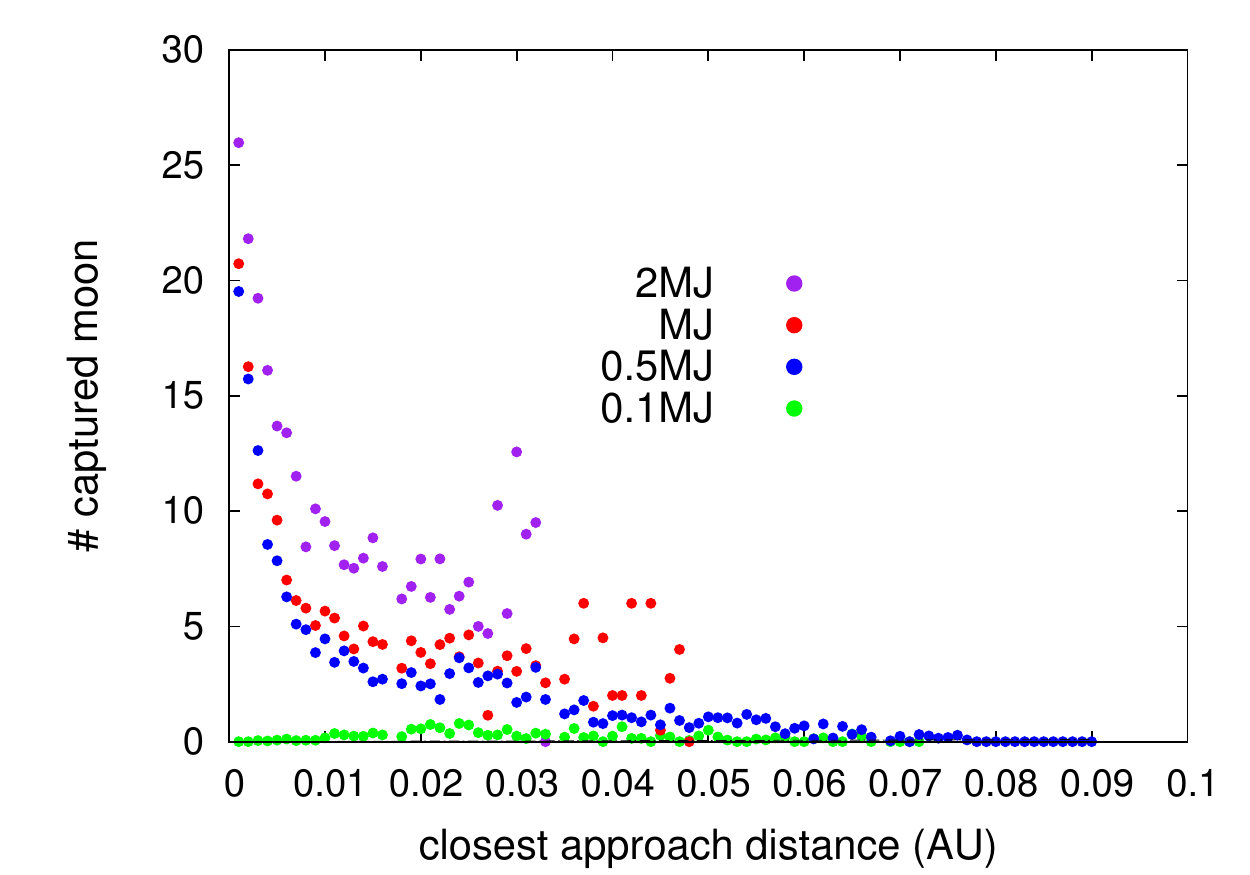}
  \caption{%
Minimum close encounter distance vs. average number of moon captures.  Figure legend as in figure \ref{figk1}.  
   }\label{figk2}
\end{figure*}

\begin{figure*}[!t]
   \centering
   \includegraphics[width=0.65\textwidth]{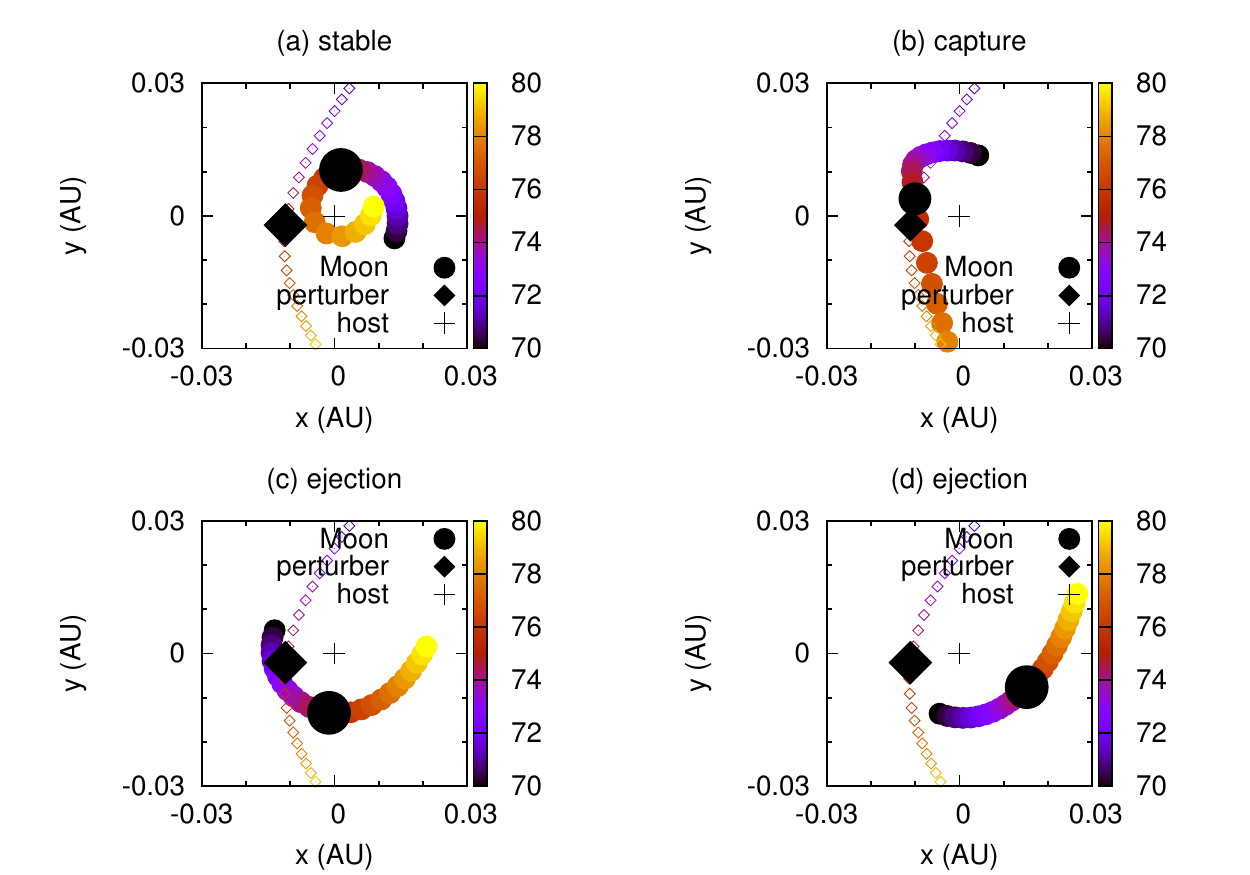}
  \caption{%
Close encounter geomentry vs. moon dynamical outcome.  The same planetary close encounter event is split into four subfigures.  Each subfigure features a different moon in the system.  In each subfigure, the host planet is represented by the cross sign at the origin, the trajectory of the perturbing planet is represented by empty circles, and the trajectory of the moon represented by empty squares.  Their trajectories change in color from purple to yellow as time evolves, as is labeled by the vertical color axis (time in days).   The solid black dots represent the location of the perturber and the moon at the time of closest approach.  See the main text for further simulation details. 
  }\label{figk3}
\end{figure*}

Regarding the moon capture rate, like the moon survival rate, the most important determining factor is $d_{min}$, and the perturber mass the second.  Figure \ref{figk2} shows the average number of moons captured by the perturber from the host planet in simulations with different $d_{min}$'s.  A perturber that has come closer to the host planet is able to pass by more moons, and because capture of moons by the perturber requires the perturber to be close to moons (fig. \ref{figk3}(b)), a smaller $d_{min}$ yields a higher
capture rate.  For planet masses ($0.5 \mbox{--} 2 M_J$) comparable to the host planet, the capture rate is roughly inversely-proportional to $d_{min}$.  Larger perturber mass slightly enhances the capture rate.  As seen from figure \ref{figk2}, the moon capture rate is only significant ($> 0.1$) for $d_{min}$ under $\sim 0.03 AU$ for simulations with $0.5\mbox{--}2 M_J$ perturbers.  0.1 $M_J$ perturbers have nearly no capability of capturing moons
from a 1 $M_J$ host planet during their encounter, as seen from the average number of captures, but in a few cases the capture rate can reach $1\mbox{--}5\%$.

Figure \ref{figk3} shows how the close encounter geometry can affect the dynamical outcome of moons.  The scenario contains an $M_J$ host planet and an $M_J$ perturber. $d_{min}$ for this encounter event is 0.01 AU, shorter than the initial semi-major axis of the four moons (0.014 AU / 0.042 $R_H$) with their angular positions evenly spaced out. Despite having the same initial orbital radius, the moons had a range of dynamical outcomes. In (a), the moon is dragged in the forward
direction and changes its orbit but stays stable.  In (b), the moon heads toward the planet and is able to reach it with the right timing and configuration from the back of the planet for a smooth transfer onto a stable orbit around the perturber to happen.
In (c) and (d), the moons are dragged in the backward direction by the planet and thus open up their orbits to become unbound.  In summary, moons on the same orbits can end up with different dynamical outcomes based on their relative distance from the perturber and also the alignment of acceleration from the perturber with the velocity vector of the moon at the time of close encounter.

\section{RESULT: FULL INTEGRATION}\label{sec:full} 

\subsection{Simulation setting}

In set 2, we simulate the dynamical evolution of exomoons in a more general context and over longer timescales.  Each simulation consists of a central star, three giant planets, and moons orbiting all the planets.  All bodies in the system are assumed to be fully formed by the time the simulation starts, and the protoplanetary disk is absent.

The star has solar mass and solar radius.  
  Each simulation includes three giant planets, and the combinations of planet masses ( chosen from 0.1, 0.3, 0.5, or 1 $M_J$) for the three planets in each simulation are sampled thoroughly and with equal frequency in the final statistics (2 simulations for systems with equal mass planets and 1 simulation for each of the other combinations of planet masses, totaling a set of 68 simulations).  Simulations where planet collisions occur are rerun, because planetary collisions could introduce complexities to moon survival.  This procedure may introduce bias into the final statistics because roughly a quarter of three-planet systems have planet collisions \citep{raymond}, and many of the planetary collisions happen early on.  This leaves the third planet $\mbox{--}$ which does not participate in the collision nor very much in the close encounters $\mbox{--}$ with a very high moon survival rate.  The innermost giant planet is placed at 5 AU from the central star, and all planets are separated by 3.5 to 4.5 mutual Hill radii, in order for instability to set in within a reasonable timescale \citep{chambers96, marzari}.   In addition, they are given initial eccentricities randomly sampled from 0.02 $\mbox{--}$ 0.1, in order to shift the first onset of instability to an earlier time.  To avoid immediate collisions between the giant planets upon orbit crossings, they are also given a small mutual inclination of 0.01$\degree$.  The giant planets are treated realistically as oblate spheroids, with their quadrupole moment $J_{2} = 0.0147$ equal to that of Jupiter, and the spin of the planet is allowed to evolve in this simulation set.

Each of the 3 giant planets in a simulation initially has 10 satellites.  Each satellite is placed at a different planetocentric orbital distance from $0.01 R_H$ to $0.35 R_H$ (0.0016 - 0.24 AU for the entire simulation set) on circular orbits.  The inner boundary of $0.01 R_H$ is close to Io's orbital distance from Jupiter, a distance far enough away from the Roche limit to avoid tidal mass loss and disintegration due to the tidal field of the planet \citep{guillochon}.  The outer boundary lies within 0.5 $R_H$, the Hill stability limit for prograde satellites \citep{domingos}.  The moons are initially on circular and nearly co-planar orbits (0 $\mbox{--}$ 0.02$\degree$) with their host planet's orbital plane and equatorial plane.  All simulations in set 2 are integrated to 1 million years. The final results discussed below are drawn from this full set of simulations unless specified.  A subset of the above simulations are further integrated up to 10 million years.  The energy error $\frac{dE}{E}$ accumulated in the simulation due to the integrator is always under $10^{-4}$, which is an adequate threshold for multi-planet systems \citep{barnes04,raymond}.   A couple of the simulations with $\frac{dE}{E} \sim 10^{-4}$ in  the above subsets are rerun or ruled out.  All simulations in set 2 are integrated to 100 million years for testing the stability of "moons" that end up orbiting the star.  This set has a higher rate of exceeding the threshold $\frac{dE}{E}$ ($\sim 20\%$) but those cases are ruled out.  All moons are massless test particles that do not perturb each other or the planets.

\subsection{Moon orbital evolution under different perturbations}

During the instability phase of the system, when planets sometimes encounter each other closely, and have strong secular interactions, moons' orbits evolve under the influence of various perturbations. The most influential and in most cases the strongest perturbation comes from the close approach of the perturbing planet to moons.  In addition, as the planets' and moons' orbits evolve through the instability phase, the stability of moons are affected.  Changes in the host planet's semi-major axis, eccentricity and inclination affect the size of the planet's Hill sphere; changes in the moon's own orbits also cause them to become Hill unstable.  A rise in the planets' and moons' inclination can put moons under Kozai-like perturbations.  The spin evolution of planets could also affect the stability of moons that are close to the critical semi-major axis, as will be further explained below.  Mutual planet secular perturbations also perturb moons.  Different types of orbital instability may also have been involved in destabilizing the moons.

\subsubsection{Direct effect of planetary close approach}

\begin{figure*}[!t]
%\vspace*{-1cm}
    \centering
   \includegraphics[width=0.95\textwidth]{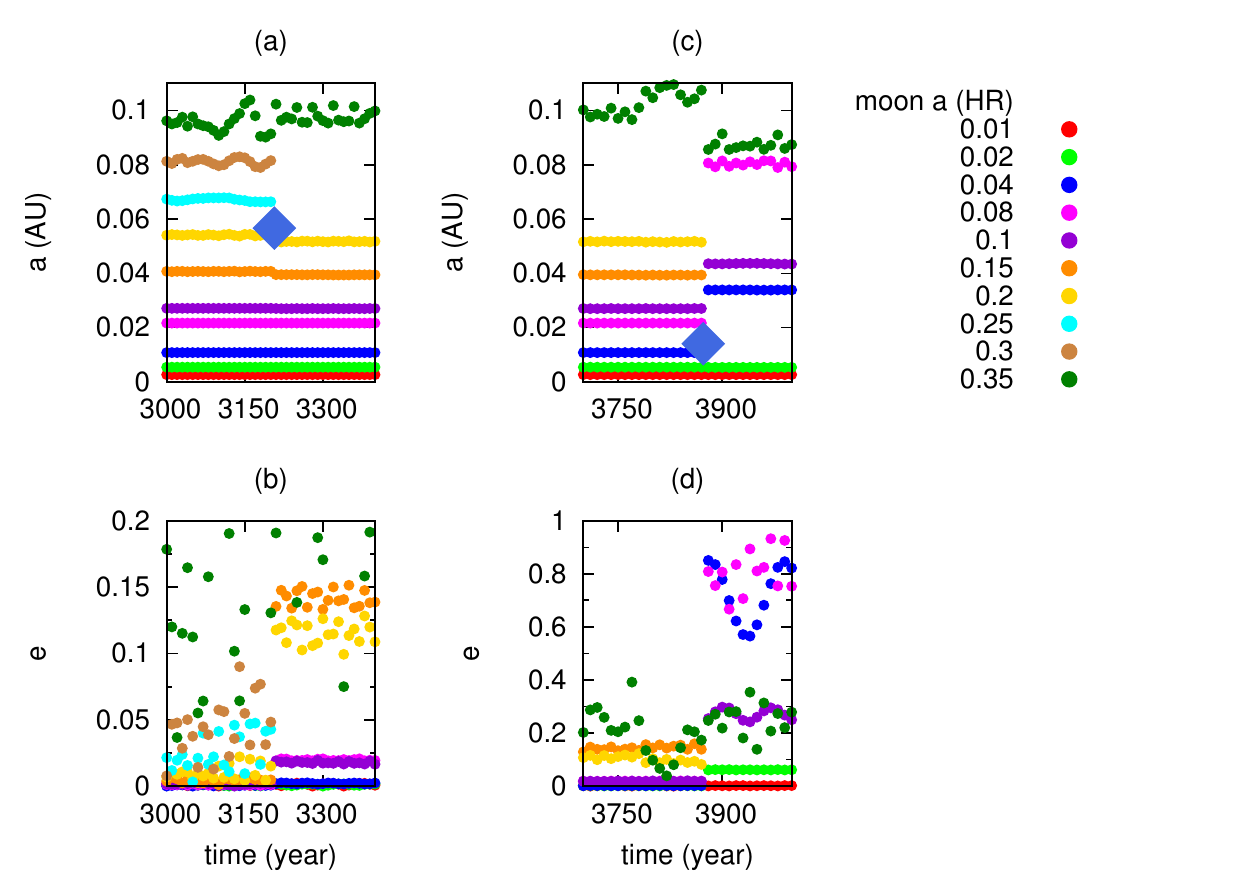}
  \caption{% 
Evolution of moon \textit{a}, \textit{e} early in two early close encounter events.  (a) and (b) features the 1st close encounter, and (c) and (d) feature the 2nd one in the same simulation.  The two events are two consecutive effective close encounters.  Each colored curve represents a moon with their initial orbital distance in $R_H$ labeled in the legend.  All moons orbit the same 0.5 $M_J$ host planet.
In (a) and (b), at $\sim$ 3200 yr, a 0.1 $M_J$ perturbing planet (blue diamond) makes a very close approach to the moon hosting planet with $d_{min} =  0.057 AU $.  (c) and (b) features a subsequent close encounter at $\sim$ 3850 yr with $d_{min} = 0.014 AU$.  
  }\label{fige1}
\end{figure*}

During a close encounter, planets and moons have strong gravitational interactions over a brief period of time, typically on the order of $10^{-1}$ to 1 years. Therefore, the time
evolution of orbital elements of planets and moons experience an instantaneous change at close encounters, as shown in fig. \ref{fige1}.  The planet's spin axis also experiences a small sudden change at a close encounter, causing a slight change in the satellite's plane of nodal precession.  But the inclination of the moon from the equatorial plane often changes more dramatically during a close encounter.  Depending on the geometry of the encounter, their semi-major axis, eccentricity and inclination may increase or decrease.  Although the close encounter geometry is random, the overall eccentricity of the moon tends to increase than to decrease.
Shown in figure \ref{fige1} is a typical case of how the moons' eccentricity evolves during the instability phase.  At a very close encounter
that is strong enough to perturb most of the moons, their eccentricities increase instantaneously by a large amount.
Subsequent milder encounters are able to perturb stable moons that are already eccentric, so then the eccentricity increases or decreases depending on the encounter geometry. Sometimes more
 encounters push the moons onto highly eccentric orbits or destabilize them.

Figure \ref{fige1} also demonstrates how a very close encounter determines the moon's orbit and dynamical outcome.
If a perturbing planet has a mass comparable to the host planet, and it approaches the host planet more closely than some of the moons, moons exterior to its closest approach distance all become destabilized.  If the perturbing planet is much less massive than the host planet (ie. at least 3 times smaller, as in fig. \ref{fige1}), not all moons become destabilized.  This demonstrates again how $d_{min}$ determines the moon stability limit as in section \ref{sec:single}, and also why close-in moons tend to have higher survival rates. In figure \ref{fige1} (a) and (b), moons interior to the closest approach distance exhibit a trend of eccentricity growth correlated to semi-major axes.  In figure \ref{fige1} (c) and (d), the close encounter further randomized the outcome of the previous one.  The 0.1 $M_J$ perturbing planet pumps up the eccentricity of the majority of moons exterior to and close to its closest approach distance, and destabilizes two of the most exterior moons.  Close encounter geometry and moon semi-major axes have similar effects as in the previous close encounter.    In between the two featured close encounter events, there are two close encounters with much larger $d_{min}$, far away from the moons, and together with the small mass of the perturbing planet, they cause little perturbation on moons.  The more distant moons, due to their closer distance to the perturbing planet, tend to become more eccentric.  In the case that not all moons are destroyed, moon survival is not as dependent on semi-major axis; the close encounter geometry (or the relative distance and velocity between the moon and the perturber) plays a more important role than in the previous case, although the strong gravitational binding of closer-in moons could still give them a somewhat better chance of survival.
If the closest approach distance is exterior to the moons, chances are higher the moons could survive but those within the sphere of the perturber's gravitational influence will become more eccentric, as a general trend, the larger their semi-major axes.  However, the encounter geomentry adds some uncertainties.  In summary, all factors of survival come into play together, but on longer timescales than shown in figure \ref{fige1}, the seemingly important role of minimum close encounter distance is smeared out.

Destabilized moons can collide with the host planet due to a small pericentric distance.  They can also collide with or become captured by the perturber.    Moons that are thrown onto hyperbolic
orbits can enter heliocentric orbits or get ejected out of the system as free-floating moons.  Ejection is by far the most common outcome.  Most of the moons that are scattered onto heliocentric orbits experience further close encounters with the giant planets, because the giant planets  themselves are on eccentric orbits.  In our 100 million year simulations, 94$\%$ of the moons on heliocentric orbits become ejected from the system to beyond 1000 AU.  Depending on the formation efficiency, free-floating "moons" can outnumber free-floating "planets" \citep{veras12}.

\subsubsection{Perturbation from planet orbital evolution}

Planetary orbits evolve chaotically as they scatter off each other and secularly interact in the planet-planet scattering phase.  Since planetary orbits set the moon stability criteria, moons are hosted in an environment constantly evolving with respect to the stability limit.  Different mechanisms leading to the destabilization of moons can be switched on besides direct perturbation in planetary close encounters.  The following subsections discuss relevant stability criteria. 

\paragraph{Hill stability}
The size of the Hill sphere depends on the planet's semi-major axis, eccentricity, and inclination \citep{domingos, donnison}.
Planets migrate in/out during the instability phase, or their distance from the star varies within an orbit due to a rise in eccentricity. When their Hill sphere shrinks through this path, moons that become exposed outside of the Hill sphere become destabilized without the direct perturbation of a close encounter.  The planet's inclination, coupled with eccentricity, also determines the Hill stability.

\paragraph{Critical semi-major axis}

The critical semi-major axis of a planet separates the region where the perturbation from planet oblateness and the central star dominates respectively.

\begin{equation}
        a_{crit} = \left(2\,J_2\,R_p^2\,a_p^3\, (1-e_p^2)^3\,\frac{m_p}{m_*}\right) ^{\frac{1}{5}},
\end{equation}
in the co-planar case, where $J_2$ is the quadrupole moment of the planet, $R_p$ the planet's radius, $a_p$ the planet's distance from the star, $m_p$ the planet's mass, and $m_*$ the star's mass \citep{kinoshita,deienno,nicholson}.   In this work, moons often have inclinations high enough for $a_{crit}$ to depend also on their inclinations. 
Well inside $a_{crit}$ the orbital angular momentum axis of the moon precesses rapidly around the planet's spin,
 and well outside $a_{crit}$ it precesses around the orbital angular momentum axis of the planet.  Moons close to $a_{crit}$
precess about local Laplace planes that lie in between the spin and orbital plane of the planet.
 As $a_{crit}$ of a planet coevolves with its orbit, the center of precession of moons switches between different planes, thus changing the moons' inclinations relative to their local Laplace planes.  Some moons wander close to $a_{crit}$ in systems with high obliquities, and could possibly be destabilized through this path \citep{tremaine09}, although planets evolving to high obliquities through planet-planet scattering also often acquire high inclinations and eccentricities, making it hard to pin down the exact cause of the moon's orbital instability.  Also, in high obliquity systems, the moons' inclination relative to the center of precession could change significantly as $a_{crit}$ changes, giving them stability or destabilizing them.  When a planet evolves to high inclinations, a path may be open to destabilizing its moons.  As planet inclination reaches $\sim 40\degree$, the Kozai mechanism starts to act on the moon that then stays close to and precesses around the orbital plane, causing the eccentricity and inclination to oscillate in a coupled manner.  

The scenarios above do not involve any direct perturbation from planetary close encounters, but are caused by evolved planetary orbits induced by close encounters plus mutual secular perturbation.  As a side note, however, only a small fraction ( $< 5 \%$) of surviving planets or planets within 100 AU from the central star in planet-planet scattering have inclinations above $40\degree$ \citep{chatterjee, raymond}, so this is not an important mechanism to destabilize the orbits of moons in the planet-planet scattering scenario.

\subsubsection{Evolved moon orbit}
Besides the evolution of planetary orbits, the evolution of the orbits of moons can also destablize or change themselves by evolving beyond the stability limit.  Oftentimes, the orbits of moons evolve when there is an effective close encounter, although mutual interactions among planets when they are outside each other's Hill sphere also affect moons.
As their orbits change, they can become exposed outside of the Hill sphere and thus become destabilized or more stable.  They can also shift across $a_{crit}$ as they radially migrate in and out, which cause changes in their inclinations relative to the center of orbital precession (local Laplace plane).  Close encounters can throw them onto random inclinations relative to their center of precession and affect their stability.

\subsubsection{Secular evolution of moon post instability}

After the planetary system has been restored to stability via removal of 1 or 2 planets, there are no more planetary close encounters.  Most of the moons simply evolve in a stable manner, with their semi-major axis staying fixed, and eccentricity and inclination undergoing sinusoidal oscillation periodically as a consequence of the precession of their orbital angular momentum around the local Laplace plane.  For some moons inside $a_{crit}$ with inclinations greater than 40$\degree$ from the equatorial plane, the inclination and eccentricity will be slowly oscillating, out of phase with each other by 180$\degree$, indicating that the Kozai mechanism is in action.  Moons with inclinations under 40$\degree$ undergo secular evolution, with a fixed semi-major axis and periodically oscillating eccentricities and inclinations.  The same argument applies to moons exterior to $a_{crit}$ inclined relative to the orbital plane.  Those that undergo Kozai evolution with high inclinations could become highly eccentric or unstable in the longer term.

\subsection{Statistics}
Table \ref{table1} summerizes possible dynamical outcomes for moons and the respective probability.  The moons that remain planet-bound may stay around their original host planet, around a new host planet that captures it away from the original, or around a planet that itself is ejected as a free-floater.  Other moons may stay on heliocentric orbits, or collide with the host planet, the perturbing planet, or the star.
Multiple close encounters throughout the planet-planet scattering phase are very efficient for removing moons.  17$\%$ of the moons in the orbital range of 0.01 $\mbox{--}$ 0.35 $R_H$ remain bound to the host planet after 1 million years of integration.  This survival rate with regard to the entire moon population signifies how much of its parameter space allows for moons to be stable.  Inferring from the above numbers, and collapsing
the parameter space to the innermost region, the moon stability limit is $\sim <$ 0.1 $R_H$.  For a similar argument from a different perspective, of the 17$\%$, 88$\%$ are initially within 0.1 $R_H$.  In other words, planet-planet scattering truncates the primordial moon disk at least as close as 0.1 $R_H$.  If considering only moons interior to 0.04 $R_H$, inside which the regular moons of the Solar System gas giant planets are located, moon survival rate amounts to $\sim 42\%$, forecasting a decent survival chance for moons on Galilean-moon like orbits.  The rest of the destabilized moons ($83\%$) are removed via various paths, as shown in table \ref{table1}.  Ejection out of the system is the most 
common outcome ($\sim 41\%$), given the small mass of moons relative to the planets (here moons are massless).
  Collision of moons with the other big bodies is also equally dominant($\sim 37\%$); among them, collision with the planets are more frequent than
with the central star due to the proximity of moons with planets, especially during planetary close encounters.  The above two paths remove the majority of the moons.  The
remaining unstable moons stay in different places in the system; $\sim 2\%$ of them are captured by the perturbing planet, and $\sim 2\%$ of them orbit planets that gets ejected out of the system as free-floaters.  Capture of moons via exhange when two planets approach each other closely is a much less efficient way of producing irregular satellites than capture from circumstellar materials probably because of the large number of the latter and that it requires a particular configuration $\mbox{--}$ small $d_{min}$.  After 1 million yr, $\sim 22\%$ of all moons initially stable around their host planets have destabilized and ended up orbiting the star; however, further integration to $10^8$ yr removed most of these heliocentric "moons", with the rate reduced to $\sim 1\%$.  The moons removed by $10^8$ yr are incorporated into the final statistics in this section.

% moon dynamical outcome

\begin{table}[]
%\captionsetup{font=scriptsize}
\centering
\begin{tabular}{|r|l|}
    \hline
    Outcome(Surviving) & probability(\%) \\
    \hline
    Host planet bound & 17 \\
    Captured by perturber & 2 \\
    Free-floater bound & 2 \\
    Heliocentric & 1 \\
    \hline \hline
    Outcome (removed)  & probability(\%) \\
    \hline
    ejected            & 41 \\
    Collision with planet & 20 \\
    Collision with star & 17 \\
  \hline
\end{tabular}
\caption{Probability of each dynamical outcome for all moons in the $10^6\mbox{--}10^8$ yr simulations in set 2.}
\label{table1}
\end{table}

$10^6$ years is 1-2 orders of magnitude shorter than required for a subset of simulations to finish the instability phase.  A sample subset of simulations taken from those that haven't ended the instability phase are further integrated up to $10^7$ yr.  The survival rate of moons in the subset decreases in half for moons across all orbital distances.  Extrapolating from the subset, the survival rate of the Galilean moons after instability ends can be around $20\mbox{--}40\%$, lower than predicted by the $10^6$ yr simulations in set 2.    The number of close encounters and the magnitude of their perturbations between $0 \mbox{--} 10^6$ yr and $10^6 \mbox{--} 10^7$ yr are comparable. Although the temporal length of the latter
makes perturbation from close encounters more diffused, the change in survival rate and how perturbed the surviving moons' orbits are not
insignificant.

\subsection{Survival factors}

\begin{figure*}[!t]
    \centering 
   \includegraphics[width=0.65\textwidth]{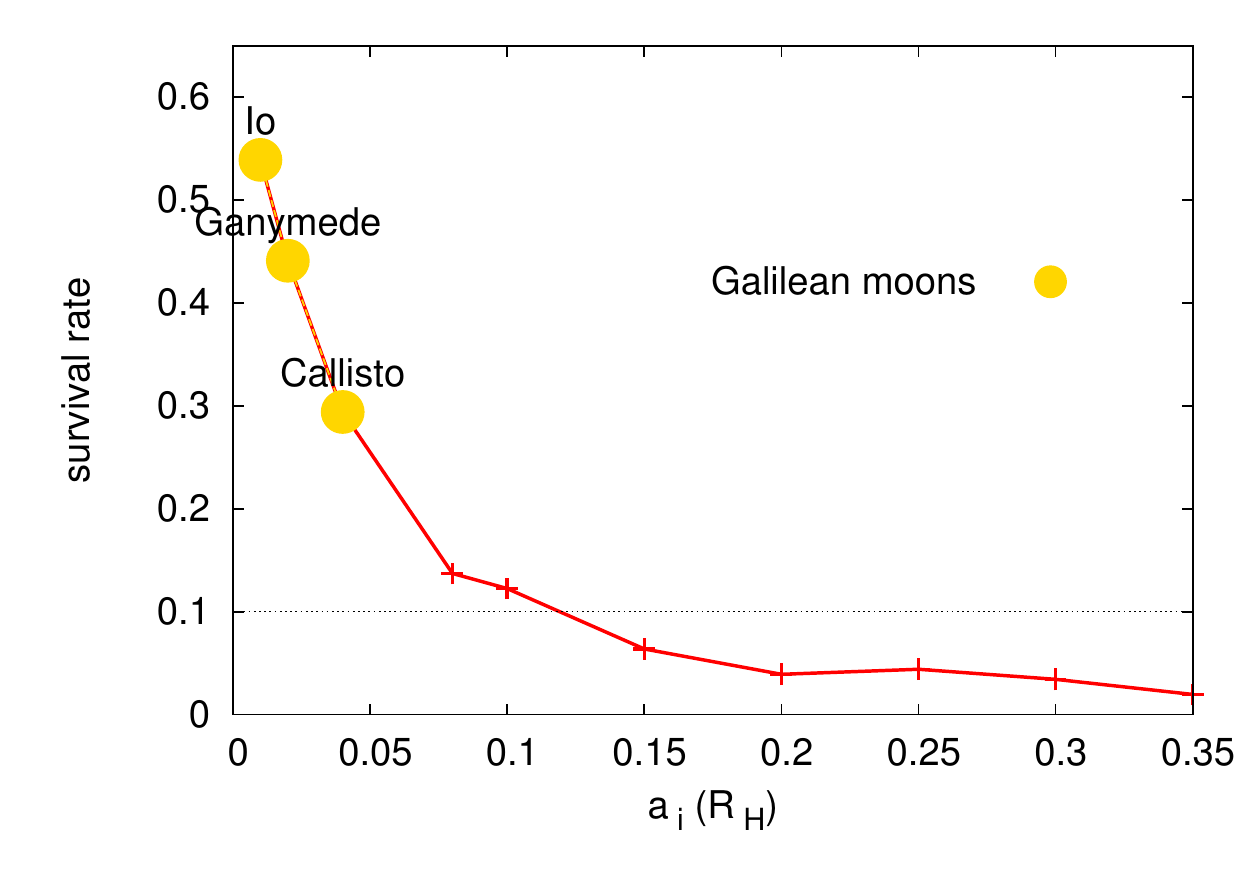}
  \caption{% 
   Moon initial semi-major axis ($a_i$ ($R_H$)) vs. moon survival rate. The moon survival rate is predetermined by the moon's initial position in the host planet's Hill sphere.  In the semi-major axis range 0.01 - 0.04 $R_H$, which is roughly where the four Galilean moons are located, survival rate drops exponentially.
  }\label{figr2}
\end{figure*}

\begin{figure*}[!t]
    \centering 
   \includegraphics[width=0.65\textwidth]{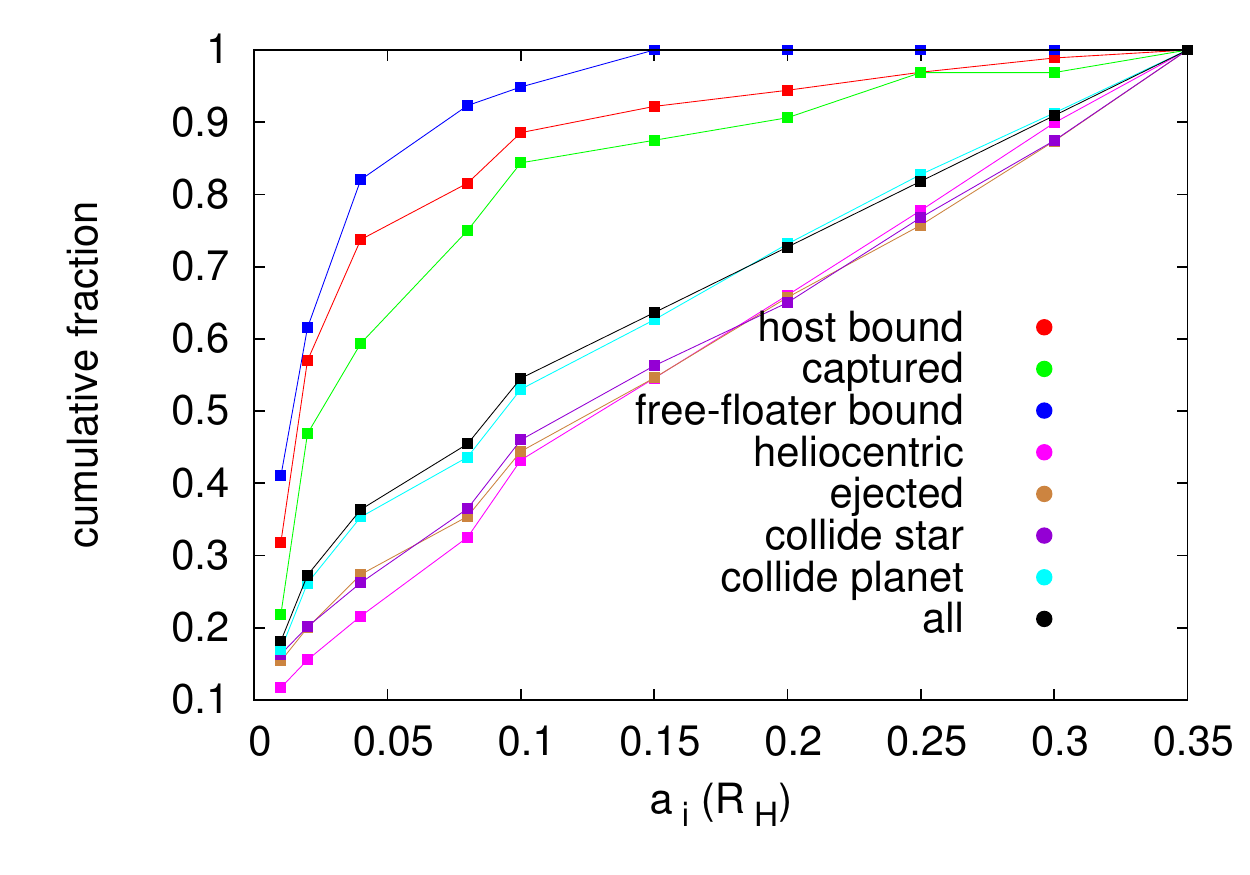}
  \caption{%
 Moon initial semi-major axis $a_i$ ($R_H$) vs. moon dynamical outcome.  Where moons will end up in planetary close encounters are also predetermined by their initial semi-major
axis, not definitely but probabilistically.  In the color legend from top to bottom are moons (1) bound to the original host planet (red), (2) captured by the perturber (green), (3) bound to the free-floating planet (blue), (4) orbiting the star (pink), (5) ejected out of the system as free-floating moons (brown), (6) colliding with the star (purple), (7) colliding with planets (light blue) , followed by the overall moon population in black.
  }\label{figr3}
\end{figure*}

\begin{figure*}[!t]
    \centering
   \includegraphics[width=0.65\textwidth]{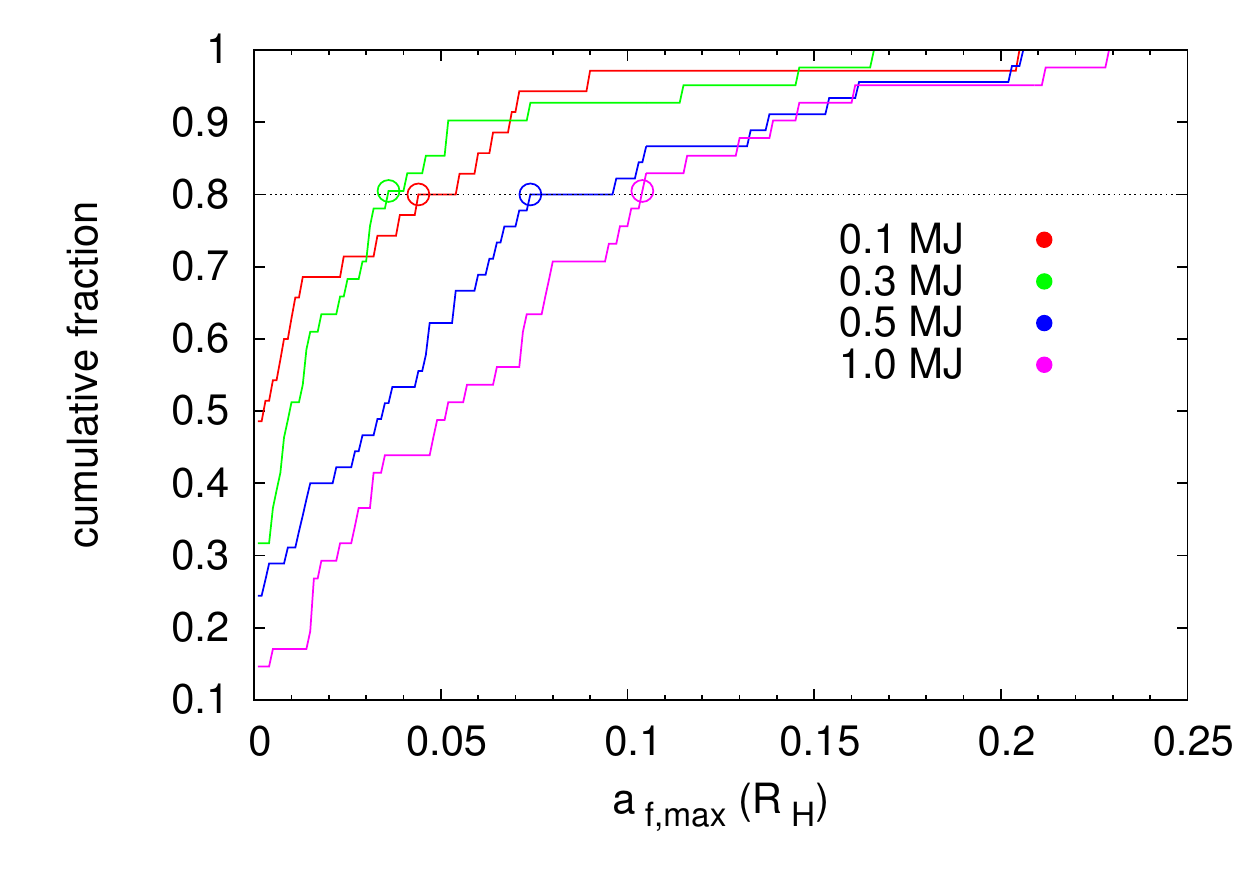}
  \caption{%
Planet mass vs. moon stability limit.
The axes of the figure represent the cumulative fraction of planets in semi-major axis of their outermost surviving moon (in $R_H$).  Different colors represent planets with different masses in $M_J$.  
  }\label{figr6}
\end{figure*}

Key factors for moon survival include various system parameters.  Parameters not observable in actual systems include the initial orbital distance of moons and properties of planetary close encounters such as closest approach distance, number of close encounters, and close encounter geometry. Observables include planet mass and planetary orbits (semi-major axis, eccentricity, inclination, and obliquity).  All of the above parameters work together in determining the dynamical outcome and stability of moons, so how moons' orbits would change can only be predicted with a single parameter in a probabilistic sense.  Moreover, the long length of the simulations and the large number of close encounters (the average number of close encounters in a $10^6$ yr simulation is 224), or the large removal rate of moons probably smears out any clear correlation between some of the parameters and the moon survival rate within the simulated set.  Figures \ref{figr2} to \ref{figr6} compare various
factors with moon survival rate and dynamical outcome.  The stability limit of moons around the planet is defined at the location of the outermost surviving moon in the Hill sphere.  Moon survival rates are nearly equivalent to the stability boundary because moons are all set at different orbital distances around each planet,
and due to strong perturbations by close encounters, survival of an outer moon often ensures that moons interior to it also survive.  However, the encounter geometry sometimes adds uncertainties to the relation between moon survival and the stability boundary.  Note that parameters related to close encounters do not correlate with moon survival rates in the full-length simulations, unlike what has been shown in single flyby events in section \ref{sec:single}.

\subsubsection{Moon initial semi-major axis}

  \paragraph{$a_i$ vs. survival rate}  
The planetocentric distance of moons determines their survival rate.  Quite intuitively, inner moons are more likely to survive because they have several advantages.  As a typical case in figure \ref{fige1}, after the perturber passed by, the inner moons usually experience less change in their orbits than do the outer moons. Closer-in moons also have a lower probability of being perturbed by an encounter since moons are not sensitive to encounters far from them.  In figure \ref{fige1}, in both of the close encounter events, most of the moons interior to the 
perturbing planet experience little perturbation while those exterior to and near it become destabilized or highly eccentric.  Different encounter geometries do add some randomness to the trend. 

As shown in figure \ref{figr2}, survival rate drops rapidly as a moon's semi-major axis increases.  Moons at Io-like distance (0.01 $R_H$) have a survival probability $\sim$0.5$\mbox{--}$0.6.  Beyond 0.1 Hill radii, the chances of survival are very low ($<$ 0.1).  The survival rate for the Galilean satellites by the end of $10^6$ yr could be higher than $\sim 0.3 \mbox{--} 0.5$ as predicted by the simulations ( the outeremost satellite, Callisto , is at $\sim 0.04$ $R_H$ from Jupiter), because in the simulations they orbit around planets of all masses but in the dynamical history of the Solar System Jupiter was the most massive planet by the time the satellites formed \citep{canup}.  Their survival rate could also be lower if the Solar System hasn't ended its instability phase by $10^6$ yr.

  \paragraph{$a_i$ vs. dynamical outcome}
The dynamical outcomes of moons are predetermined in a probabilistic sense by their initial location within the Hill sphere.  Figure \ref{figr3} shows the cumulative distribution in the initial semi-major axis of moons that have different dynamical outcomes.  The overall population (black line) divides the planet-bound moons and the unbound moons.  Not surprisingly, the planet-bound moons tend to occupy the inner part of the Hill sphere, with their curves consistently lying on the left divison of the overall population.  Among them, those that stay stable around an ejected free-floating host planet lie in the most interior region, likely due to the fact that they need a sufficiently strong gravitational binding and enough separation with the perturber to survive the planet-ejecting close encounters.  By contrast, moons that become heliocentric, including those that later collide with or orbit the star and those that are ejected out of the system, lie consistently on the right hand side of the overall population, and the three curves are very close to each other, since right after the close encounter that destabilize them, they share the same outcome.

\subsubsection{Planet mass}
The mass of the host planet determines the extent of its moons' gravitational binding, which if large compared to perturbations, the moons are more likely to survive.  Therefore the more massive the planet, the more stable its moons
usually are.  Figure \ref{figr6} shows the cumulative distribution of the outermost surviving moon in semi-major axis on planets with different masses. 
Moons around more massive planets have wider distributions across the Hill sphere; in other words,
 they are more likely to have a larger stability boundary.  To have moons surviving beyond 0.1 $R_H$, the probability for planets with $M_J$, 0.5 $M_J$, 0.3 $M_J$, and 0.1 $M_J$ are $\sim 35\%$, $\sim 20\%$, $\sim 10\%$, and $\sim 5\%$ respectively.  Around 0.5 and 1 $M_J$ planets, the 80th percentile of the outermost stable moons reaches 0.1 $R_H$, whereas around 0.1 and 0.3 $M_J$ planets they are under a much tighter limit of 0.04 $R_H$.  All planet masses from 0.1 to 1 $M_J$ permit moons on Galilean satellite orbits to have good survival rates.

\subsection{Final orbits of moons}

 Moons with different dynamical outcomes have experienced different amount of perturbations, so they occupy different parts of the orbital parameter space; in other words, they are dynamically distinguishable.  As a general trend, the more they deviate from their initial orbits, the more they have been perturbed.

\subsubsection{$\Delta a$} 

Planet-bound moons have low tolerance for radial migration.  A large amount of change in the moons' semi-major axis often destabilizes it.  Moons that stay bound to stable host planets experience no dramatic changes in semi-major axis. 80\% of them migrated outward / inward by less than 46$\%$ / 32$\%$ in semi-major axis (or \%90 by less than 73\%/45\%).  Moons stably orbiting ejected free-floating planets (2$\%$) are the most tightly clustered at small semi-major axis  (in AU).  The 80th percentile reaches as far as $\sim$ 0.01 AU.  Their small orbital distance is very likely a result of the same processes identified in the previous section.  In comparison, moons that have ended up on heliocentric orbits have a wider distribution in semi-major axis than the surviving planets.

\subsubsection{Eccentricity}
Quite intuitively, moons that stay stable around their original host planet have less excited orbits than those that are destabilized.  Moons that remain stable around bound or free-floating planets have lower average eccentricities, and those captured by another planet or go into heliocentric orbits are much more eccentric.

\begin{figure*}[!t]
    \centering
   \includegraphics[width=0.65\textwidth]{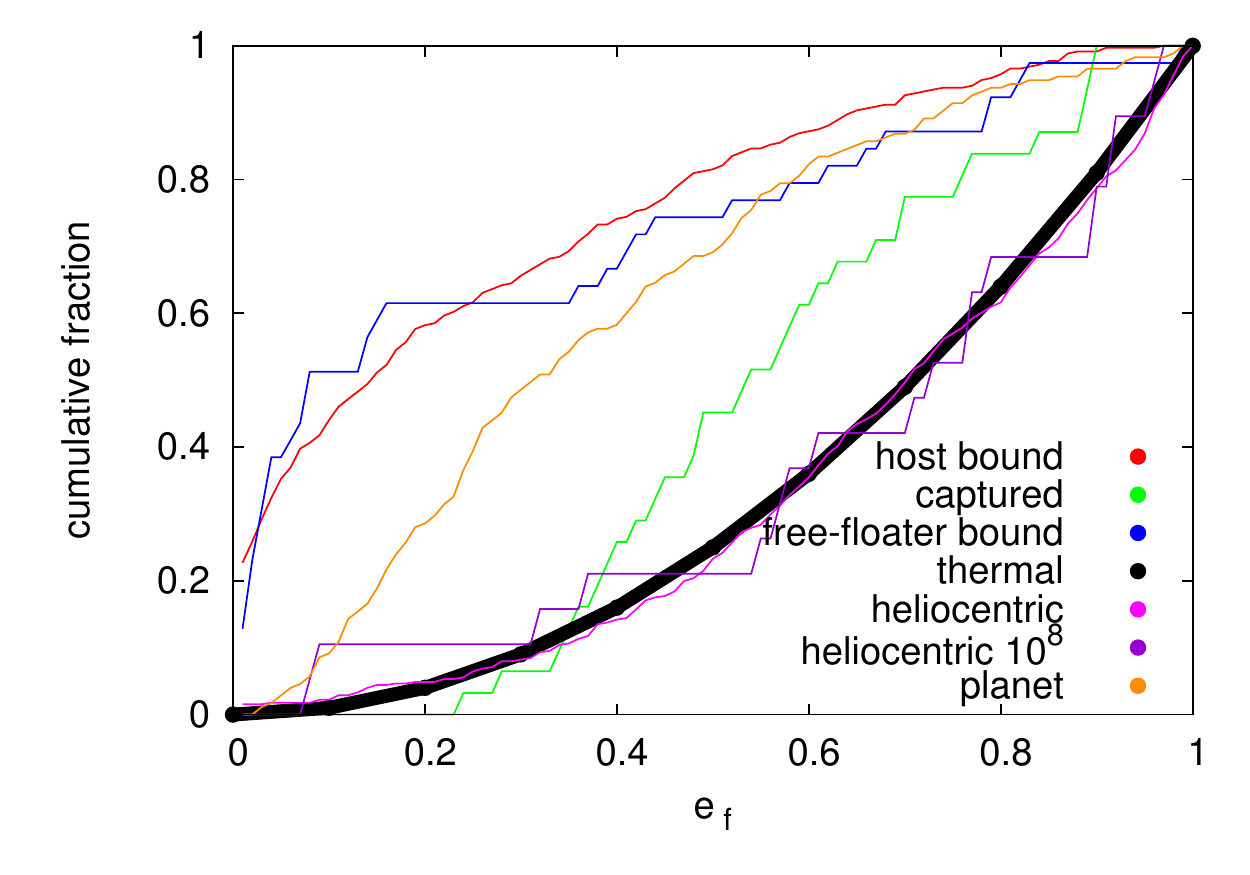}
  \caption{%
Cumulative fraction of final eccentricity of stable planets and moons with different dynamical outcomes.  In the legend different colors represent different poulations of moons.
For planet$\mbox{--}$bound moons, those bound to the original host planet are represented in red, those captured by the perturber in green, and those bound to free-floaters in blue.  Moons on heliocentric orbits by the end of $10^6 / 10^8$ years are represented in pink / purple.  Planets surviving around the central star are represented in black.
  }\label{figr8}
\end{figure*}

Figure \ref{figr8} shows the cumulative fraction in eccentricity of moons of different dynamical outcomes.  Primordial moons bound to the host planets are, to no surprise, the least eccentric population, since their orbits receive the least perturbation and change the least. (fig. \ref{figr8} )  44\% of them have eccentricities under 0.01, and the average eccentricity of the perturbed ones is 0.31.  Moons become eccentric
once an encounter strong enough to perturb them occurs, and the probability to recover via subsequent encounters is extremely low; more violent encounters usually only 
increase eccentricity by a large amount, and milder encounters have a chance to damp down the eccentricity but only by a moderate amount.
The average eccentricity of moons bound to free-floating planets is $\sim$ 0.32, higher than moons around stable planets, and the 80th percentile reaches e $\sim$ 0.6; as a comparison, the 80th percentile of primordial moons around stable planets reaches e $\sim 0.47$.  Moons that are captured away by a different planet have a higher average eccentricity of 0.57, and they are more evenly distributed across the eccentricity spectrum except at the low end with e $<$ 0.2, because their orbits are randomized by the capture process.  

Moons stripped away from the planet to orbit the central star have the highest average eccentricity of 0.68, the 80th percentile reaches as high as e = $\sim$ 0.91, much
higher than all the planet-bound populations, due to the higher tolerance for eccentricity in the heliocentric system than in the planetocentric system.  Both for simulations lasting $10^6$ and $10^8$ yr, all the heliocentric moons in the set form a thermal eccentricity distribution (fig. \ref{figr8}),

\begin{equation}
f(e) = 2e.
\end{equation}
 This distribution resembles that of wide binary stars and suggests that they have been sufficiently pulled and kicked around for the energy to approximate an equilibrium state \citep{jeans,kouwenhoven}.  However, as a consequence of extremely high eccentricities and the fact that the planets are also on eccentric orbits, interactions are often strong and frequent, making heliocentric moons extremely vulnerable to subsequent removal.  By the end of $10^{8}$ years, only $\sim 6\%$ of them remain in the system ($a < 1000 AU$).  However, if a planetesimal belt exists at the edge of the planetary system, it could help trap moons on the way to ejection, turning them into KBO-like objects \citep{raymond09}.  Another possibility to produce KBO-like objects out of ejected moons is that, if some instabilities happen before the dissipation of the gas disk, gas drag could act to trap a heliocentric moon by lifting its pericenter from the grasp of the perturbing planets \citep{raymond17}.  Of the majority of moons venturing beyond 1000 $AU$ (classified as "ejected" in this work), $\sim 35\%$ are able to reach at least 5000 AU with eccentricities $\sim< 1$ within $10^8$ yr.  Those moons might have a chance to damp down their eccentricities by the galactic tide and become Oort-cloud-like objects \citep{morbidelli,tremaine93}.    $20\%$ of the test simulations for Oort-cloud like candidates with fractional angular momentum change larger than $10^{-5}$ are ruled out because most of the moons that reach the orbital distance where the galactic tides becomes effective have eccentricities $\sim < 10^{-5}$ below 1.

\subsubsection{Inclination}

\begin{figure*}[!t]
    \centering
   \includegraphics[width=0.65\textwidth]{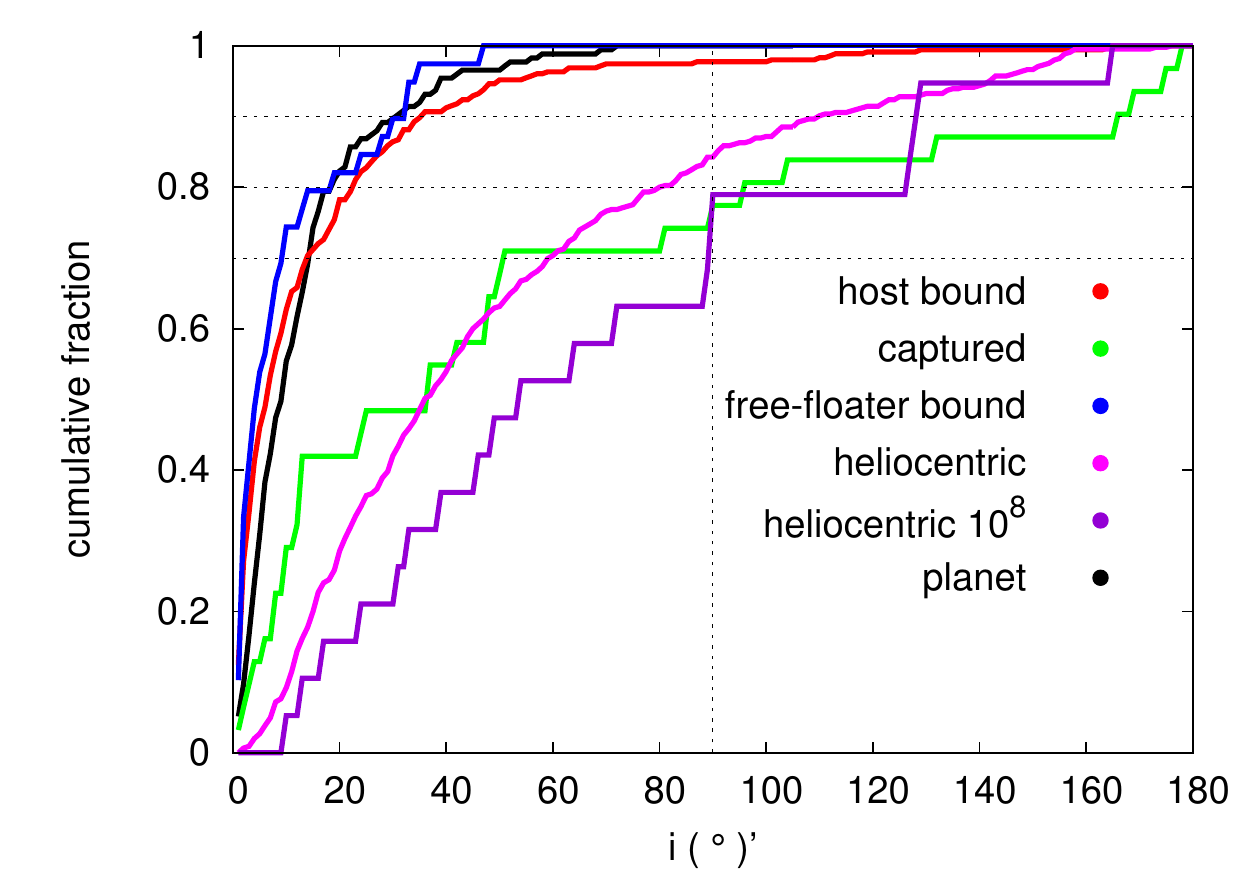}
  \caption{%
Cumulative distribution of final inclination of stable planets and moons with different dynamical outcomes.
The figure legend is identical to fig. \ref{figr8}. Inclinations are measured relative to the initial zero inclination plane of the system, if not specified.  
  }\label{figr9}
\end{figure*}

Like eccentricity, inclination is an indicator of the dynamical history of moons.  Figure \ref{figr9} shows the cumulative fraction of moons in inclination of different dynamical outcomes.  The stable populations, the host-planet-bound and free-floater-bound moons and the planets are again the least evolved in \textit{i}, as in \textit{a} and \textit{e}.  The more unstable moon populations such as the heliocentric moons and the captured ones have larger inclinations; 45$\%$ of them are above 40$\degree$ as seen from the planet's orbital plane.  And a much larger fraction of them are retrograde than other populations; 23$\%$ for moons captured by the perturbing planet, and 16$\%$ for heliocentric moons by $10^6$ yr.  In comparison, only $\sim 1\%$ of the planet$\mbox{--}$bound moons are on retrograde orbits.

Surviving moons have relatively quiet orbits, although still significantly more perturbed than the Solar System regular moons.
86$\%$ of stable primordial moons within $a_{crit}$ are under $\sim 20\degree$ as seen from the host planet's equatorial plane and 73$\%$ are above 1$\degree$, as for those exterior to $a_{crit}$, 81$\%$ are under 20$\degree$ from the host planet's orbital plane.  The inner population is significantly less perturbed than the outer one.  As for moons stable around ejected free-floating planets, 80$\%$ are under 20$\degree$ from the initial reference plane, but the number of such systems is small, making statistical interpretation difficult.  Both of the primordial populations have under $2\%$ on retrograde orbits.

Moons captured by the perturber are more evenly distributed across the inclination spectrum, including a significant portion on the retrograde side, and in general they have higher inclinations than moons bound to their original host planet; this tendency is similar to the simulated capture of satellites from planetesimals \citep{nesvorny}.
  
\section{Occurrence of free floating exomoons}

 Here we calculate the abundance of free-floating exomoons and estimate its potential contribution to the vast number of free-floating objects estimated from microlensing observations, since planet-planet scattering appears to be very efficient at ejecting moons.  \citet{sumi} obtained an observed frequency of free-floating planets per main-sequence star $\frac{N_{ff}}{N_{star}} = 1.8^{+1.7}_{-0.8}$, and \citet{mroz} found $\frac{N_{ff}}{N_{star}} = 0.25$ for Jupiter-mass free-floaters.  The equation below adopts a smilar method to \citet{veras12} to estimate the number of free-floating moons:

 \begin{equation} 
%\begin{split}
   \frac{N_{moon,ff}}{N_{star}} = f_{gp} \times f_{system, unstable} \times  n_{gp,eject}  \times n_{moon,eject}
 %\end{split}
 \end{equation}

, where $N_{star}$ is the total number of stars, $f_{gp}$ the fraction of stars with giant planets, $f_{system, unstable}$ the fraction of planetary systems that go unstable, $n_{gp,eject}$ the averaged number of ejected giant planets per system, $n_{moon}$ the averaged number of moons per system.  

The giant planet occurrence rate $f_{gp}$ has been measured by radial velocity surveys to be $\sim$0.1-0.2 for Sun-like stars \citep{cumming,mayor,rowan16} but with a stellar-mass dependence such that low-mass stars are deficient in gas giants \citep{johnson,lovis,winn, wittenmyer}.  Given the predominance of low-mass stars by number, we expect the average $f_{gp}$ to be in the $1\mbox{--}10\%$ range.  The number of giant planets that participate in a given instability $n_{gp,unstable}$, must be at least two, and may be larger.  We assume an average of 2$\mbox{--}$4.
 To match the observed eccentricity distribution of exoplanets, the fraction of unstable planetary systems is very high.  $f_{system, unstable}$ is at least 75$\%$ but is more likely more than 90$\%$ \citep{raymond,raymond11}.  $n_{moon,eject}$ equals the number of moons per system $\times$ ejection rate. The former is unknown, and we assume a value of $1\mbox{--}5$; the latter is roughly 40$\%$ from the result of simulation set 2.  Therefore, the occurence of moons per star is $\mathcal{O}(0.01\mbox{--}1)$, which predicts a galactic population of free-floating (former) moons that may be as abundant as stars.  WFIRST can detect such objects down to $\sim0.1$ Earth masses \citep{spergel}.

  Another interesting population of moons that sit stably on ejected planets has an occurence probability of order 1$\%$ per system based on the theoretical result in this work.  Replacing $n_{moon,eject}$ with the fraction of free-floating giant planets that carry moons ($\sim0.08$) and using the same estimate for other parameters as above yields an occurence rate of $\mathcal{O}(10^{-3}\mbox{--}10^{-2})$ per star.  If using the observed frequency of free-floating planets per star (0.25) \citep{mroz} and multiplying it with the fraction of moon-bearing free-floating planets from the simulations ($\sim0.08$), the occurence of moon-bearing free-floating planets per star is $\mathcal{O}(10^{-2})$.  Both approaches yields probabilities that are not insignificant.

\section{Conclusion}

We have directly simulated the survival of exomoons during giant planet scattering.  To summerize, most moons are unstable during planet-planet scattering and they have rich dynamical outcomes.  Planet-planet scattering is a very efficient way to destabilizing moons, mainly because the close approach of the relatively massive perturbing planet compared to moons, and secondly because planets themselves also experience strong mutual perturbations induced by the instability even outside of close encounters.  Only $\sim10-20\%$ of the moons within 0.35 $R_H$ remain bound to the original host planet; in other words, only $\sim10-20\%$ of the orbital parameter space in the Hill sphere allows for moons to be stable.  Close-in moons on Galilean-moon like orbits (0.01 - 0.04 $R_H$) have at least twice higher survival rates than the whole population.  The majority of moons ($\sim 80-90\%$) collide with big bodies or become ejected out of the system.  Given the high probability of ejection, there is likely a population of free-floating objects in interstellar space that were born as moons of giant planets, with an occurrence rate of $\mathcal{O}(0.01\mbox{--}1)$ per star.   Future microlensing detection is capable of observing such objects down to $\sim0.1$ Earth masses.   A very tiny fraction of moons live on interesting orbits, such as around free-floating planets (occurrence rate $\mathcal{O}(10^{-3}\mbox{--}10^{-2})$ per star), around a perturbing planet as a captured object, or on heliocentric orbits.  As moons around free-floating planets tend to be very close-in and eccentric in the planet-planet scattering scenario, there may be a good chance of having tidally heated Io's around free-floating planets.  The simulated number of moons are not abundant enough to explain the efficiency of producing Trojan-like objects from moon scattering, but intuition and the simulation results for moons on heliocentric orbits suggest that such objects can have a hard time surviving the instability phase. Since planet-planet scattering tends to destroy already-formed terrestrial planets \citep{veras05, matsumura, carrera} or their building blocks \citep{raymond11,raymond12}, a possible way to form terrestrial planets could be from the heliocentric moons remaining in the system after planet-planet scattering, although the simulation results do not predict it as a highly efficient mechanism (probability $\sim0.01$).

Properties of the close encounters, planets, and moons all come into play together in determining the stability and dynamical outcomes of moons.  Among all factors, the planet mass and the initial orbits of moons stand out as clear predictors of a moon's final outcome, though only in a probabilistic sense.  The source region of moons with different dynamical outcomes can be anticipated to some degree.  

As to the final orbits of moons, stable moons experience limited radial migration, so it is not a common phenomenon for moons to swap their orbits.  However, moons do move in and out of the planet's Hill stability limit and the critical semi-major axis during the instability.  For the Solar System gas giant planets, moons interior to the critical semi-major axis are co-planar with the equator and have tiny eccentricities \citep{deienno}, whereas moons exterior to it are inclined and eccentric \cite{jewitt07}.  In contrast, in the bulk simulated extra-solar systems, moons interior to the critical semi-major axis have a tiny probability  $\sim < 0.1$ of having unperturbed orbits like in the Solar System, giving a hint that Solar System gas giant planets might not have experienced planetary close encounters, or that they experienced very few and very mild close encounters.  In the region exterior to the critical semi-major axis and interior to the Hill stability limit for prograde moons, moons have a negligible probability $\sim 0.01$  of having unperturbed orbits, just like in the Solar System.  The Solar System irregular moons on prograde orbits could have mixed origins of primordial populations and external capture from circumplanetary materials.  However, tests in the planet-planet scattering scenario show that moons on retrograde orbits are not produced very efficiently, therefore supporting the origin of retrograde irregular satellites via other mechanisms such as capture from circumplanetary materials \citep{nesvorny}.

\section{acknowledgements}

We thank John Chambers, Tom Quinn, and John Armstrong for helpful discussions.  We are grateful to an anonymous referee for a constructive review.  This work was supported by the JWST (NASA grant NNX12AK016) and CASSINI (subcontract JPL 64969) projects.  S.~N.~R. thanks the Agence Nationale pour la Recherche for funding via grant ANR-13-BS05-0003-002 (MOJO) and NASA Astrobiology Institute's Virtual Planetary Laboratory Lead Team, funded via the NASA Astrobiology Institute under solicitation NNH12ZDA002C and cooperative agreement no. NNA13AA93A. 

\vspace*{3\baselineskip}

\vspace*{3\baselineskip}

%\afterpage{

%\afterpage{

%\afterpage{

%\afterpage{

% initial moon a
%\afterpage{

%\afterpage{

%\afterpage{

%\afterpage{

%\afterpage{


\begin{thebibliography}
\expandafter\ifx\csname natexlab\endcsname\relax\def\natexlab#1{#1}\fi


\bibitem[{Adams} \& {Laughlin} (2003)]{adams}
{Adams}, F.~C. and {Laughlin}, G. 2003, \icarus, 163, 290

\bibitem[{{Agol} {et~al.}(2005){Agol}, {Steffen}, {Sari}, \& {Clarkson}}]{agol}
{Agol}, E., {Steffen}, J., {Sari}, R., {et~al.} 2005, \mnras, 359, 567

\bibitem[{{Barnes} \& {O'Brien}(2002)}]{barnes}
{Barnes}, J.~W., \& {O'Brien}, D.~P. 2002, \apj, 575, 1087

\bibitem[{Barnes} \& {Quinn} (2004)]{barnes04}
{Barnes}, R., \& {Quinn}, T. 2004, \apj, 611, 1

\bibitem[{{Bennett} \& {Rhie}(1996)}]{bennett96}
{Bennett}, D.~P., \& {Rhie}, S.~H. 1996, \apj, 472, 660

\bibitem[{{Bennett} \& {Rhie}(2002)}]{bennett02}
{Bennett}, D.~P., \& {Rhie}, S.~H. 2002, \apj, 574, 985

\bibitem[{{Bennett} {et~al.}(2014){Bennett}, {Batista}, {Bond}, {Bennett},
  {Suzuki}, {Beaulieu}, {Udalski}, {Donatowicz}, {Bozza}, {Abe}, {Botzler},
  {Freeman}, {Fukunaga}, {Fukui}, {Itow}, {Koshimoto}, {Ling}, {Masuda},
  {Matsubara}, {Muraki}, {Namba}, {Ohnishi}, {Rattenbury}, {Saito}, {Sullivan},
  {Sumi}, {Sweatman}, {Tristram}, {Tsurumi}, {Wada}, {Yock}, {The MOA
  Collaboration}, {Albrow}, {Bachelet}, {Brillant}, {Caldwell}, {Cassan},
  {Cole}, {Corrales}, {Coutures}, {Dieters}, {Dominis Prester}, {Fouqu{\'e}},
  {Greenhill}, {Horne}, {Koo}, {Kubas}, {Marquette}, {Martin}, {Menzies},
  {Sahu}, {Wambsganss}, {Williams}, {Zub}, {The PLANET Collaboration}, {Choi},
  {DePoy}, {Dong}, {Gaudi}, {Gould}, {Han}, {Henderson}, {McGregor}, {Lee},
  {Pogge}, {Shin}, {Yee}, {The {$\mu$}FUN Collaboration}, {Szyma{\'n}ski},
  {Skowron}, {Poleski}, {Koz{\l}owski}, {Wyrzykowski}, {Kubiak},
  {Pietrukowicz}, {Pietrzy{\'n}ski}, {Soszy{\'n}ski}, {Ulaczyk}, {The OGLE
  Collaboration}, {Tsapras}, {Street}, {Dominik}, {Bramich}, {Browne},
  {Hundertmark}, {Kains}, {Snodgrass}, {Steele}, {The RoboNet Collaboration},
  {Dekany}, {Gonzalez}, {Heyrovsk{\'y}}, {Kandori}, {Kerins}, {Lucas},
  {Minniti}, {Nagayama}, {Rejkuba}, {Robin}, \& {Saito}}]{bennett14}
{Bennett}, D.~P., {Batista}, V., {Bond}, I.~A., {et~al.} 2014, \apj, 785, 155

\bibitem[{Bills} (1990)]{bills90}
{Bills}, B.~G. 1990, \jgr, 95, 14137

\bibitem[{Bills} (2005)]{bills05}
{Bills}, B.~G. 2005, \icarus, 175, 233

\bibitem[{Canup} \& {Ward} (2008)]{canup}
{Canup}, R.~M., \& {Ward}, W.~R. 2008, Europa, University of Arizona Press

\bibitem[{Carrera} {et~al.} (2016)]{carrera}
{Carrera}, D., {Davies}, M.~B., \& {Johansen}, A. 2016, \mnras, 463, 3226

\bibitem[{{Chambers}(1996)}]{chambers96}
{Chambers}, J.~E., {Wetherill}, G.~W., \& {Boss}, A.~P. 1996, \icarus, 119, 261

\bibitem[{{Chambers}(1999)}]{chambers}
{Chambers}, J.~E. 1999, \mnras, 304, 793

\bibitem[{Chatterjee} {et~al.} (2008)]{chatterjee}
{Chatterjee}, S., {Ford}, E.~B., {Matsumura}, S., {et~al.} 2008, \apj, 686, 580

\bibitem[{Cumming} {et~al.} (2008)]{cumming}
{Cumming}, A., {Butler}, R.~P., {Marcy}, G.~W., {et~al.} 2008,
\pasp, 120, 531

\bibitem[{Deienno} {et~al.} (2011)]{deienno}
{Deienno}, R., {Yokoyama}, T., {Nogueira}, E.~C., {et~al.} 2011, \aap, 536, A57

\bibitem[{Dobos} {et~al.} (2017) {Dobos}, {Heller}, \& {Turner}]{dobos}
{Dobos}, V., {Heller}, R., \& {Turner}, E.~L. 2017, \aap, 601, 91

\bibitem[{{Domingos} {et~al.}(2006){Domingos}, {Winter}, \&
  {Yokoyama}}]{domingos}
{Domingos}, R.~C., {Winter}, O.~C., \& {Yokoyama}, T. 2006, \mnras, 373, 1227

\bibitem[{{Donnison}(2010)}]{donnison}
{Donnison}, J.~R. 2010, \mnras, 406, 1918

\bibitem[{Ford} \& {Rasio} (2008)]{ford}
{Ford}, E.~B., {Rasio}, F.~A. 2008, \apj, 686, 621

\bibitem[{Forgan} \& {Kipping} (2013)]{forgan}
{Forgan}, D., {Kipping}, D. 2013, \mnras, 432, 2994

\bibitem[{{Frouard} \& {Yokoyama}(2013)}]{frouard}
{Frouard}, J., \& {Yokoyama}, T. 2013, Celestial Mechanics and Dynamical
  Astronomy, 115, 59

\bibitem[{{Gong} {et~al.}(2013){Gong}, {Zhou}, {Xie}, \& {Wu}}]{gong}
{Gong}, Y.-X., {Zhou}, J.-L., {Xie}, J.-W.,{et~al.} 2013, \apjl, 769,
  L14

\bibitem[{{Guillochon} {et~al.}(2011){Guillochon}, {Ramirez-Ruiz}, \&
  {Lin}}]{guillochon}
{Guillochon}, J., {Ramirez-Ruiz}, E., \& {Lin}, D. 2011, \apj, 732, 74

\bibitem[{{Han}(2008)}]{han08}
{Han}, C. 2008, \apj, 684, 684

\bibitem[{{Han} \& {Han}(2002)}]{han02}
{Han}, C., \& {Han}, W. 2002, \apj, 580, 490

\bibitem[{{Heller}(2012)}]{heller12}
{Heller}, R. 2012, \aap, 545, L8

\bibitem[{{Heller} \& {Zuluaga}(2013)}]{heller13a}
{Heller}, R., \& {Zuluaga}, J.~I. 2013, \apjl, 776, L33

\bibitem[{{Heller} \& {Barnes}(2013)}]{heller13b}
{Heller}, R., \& {Barnes}, R. 2013, Astrobiology, 13, 18

\bibitem[{Heller} {et~al.}(2014)]{heller14c} 
{Heller}, R., {Williams}, D., {Kipping}, D., {et~al.} 2014, Astrobiology, 14, 798 

\bibitem[{Heller} {et~al.} (2014)]{heller14a}
{Heller}, R. \& {Barnes}, R. 2014, Int. J. AsBio, 14, 335

\bibitem[{Heller} {et~al.} (2014)]{heller14b}
{Heller}, R. 2014, \apj, 787, 14

\bibitem[{Hilton} (1991)]{hilton91}
{Hilton}, J.~L. 1991, \aj, 102, 1510

\bibitem[{Hinkel} \& {Kane} (2013) ]{hinkel}
{Hinkel}, N.~R., {Kane}, S.~R. 2013, \apj, 774, 27

\bibitem[{{Holman} \& {Murray}(2005)}]{holman}
{Holman}, M.~J., \& {Murray}, N.~W. 2005, Science, 307, 1288

\bibitem[{{Holman} \& {Wiegert}(1999)}]{holman99}
{Holman}, M.~J., \& {Wiegert}, P.~A. 1999, \aj, 117, 621

\bibitem[{Hong} {et~al.} (2015)]{hong15}
{Hong}, Y.-C., {Tiscareno}, M.~S., {Nicholson}, P.~D., {et~al.} 2015, \mnras, 449, 828

\bibitem[{Jeans} (1919)]{jeans}
{Jeans}, J.~H. 1919, \mnras, 79, 408

\bibitem[{Jewitt} \& {Haghighipour} (2007)]{jewitt07} 
{Jewitt}, D., \& Haghighipour, N.\ 2007, \araa, 45, 261 

\bibitem[{Johnson} {et~al.} (2007)]{johnson}
{Johnson}, J.~A., {Butler}, R.~P., {Marcy}, G.~W., {et~al.}, 2007,
\apj, 670, 833

\bibitem[{Juri{\'c}} \& {Tremaine} (2008)]{juric}
{Juri{\'c}}, M. , \& {Tremaine}, S. 2008, \apj, 686, 603

\bibitem[{{Kaltenegger}(2010)}]{kaltenegger}
{Kaltenegger}, L. 2010, \apjl, 712, L125

\bibitem[{Kane} (2017)]{kane}
{Kane}, S.~R. 2017, \apjl, 839, L19

\bibitem[{{Kasting} {et~al.}(1993){Kasting}, {Whitmire}, \&
  {Reynolds}}]{kasting}
{Kasting}, J.~F., {Whitmire}, D.~P., \& {Reynolds}, R.~T. 1993, \icarus, 101,
  108

\bibitem[{Kinoshita} \& {Nakai} (1991)]{kinoshita}
{Kinoshita}, H. \& {Nakai}, H. 1991, CeMDA, 52, 293

\bibitem[{{Kipping}(2009)}]{kipping09a}
{Kipping}, D.~M. 2009, \mnras, 392, 181

\bibitem[{{Kipping} {et~al.}(2012){Kipping}, {Bakos}, {Buchhave},
  {Nesvorn{\'y}}, \& {Schmitt}}]{kipping12}
{Kipping}, D.~M., {Bakos}, G.~{\'A}., {Buchhave}, L., {et~al.} 2012, \apj, 750, 115

\bibitem[{{Kipping} {et~al.}(2009){Kipping}, {Fossey}, \&
  {Campanella}}]{kipping09c}
{Kipping}, D.~M., {Fossey}, S.~J., \& {Campanella}, G. 2009, \mnras, 400, 398

\bibitem[{Kipping} {et~al.} (2015)]{kipping15}
{Kipping}, D.~M., {Schmitt}, A.~R., {Huang}, X., {et~al.} 2015, \apj, 813, 14



\bibitem[{Kouwenhoven} {et.~al} (2010)]{kouwenhoven}
{Kouwenhoven}, M.~B.~N., {Goodwin}, S.~P., {Parker}, R.~J., {et~al.} 2010, \mnras, 404, 4

\bibitem[{Lin} \& {Ida} (1997)]{lin}
{Lin}, D.~N.~C., \& {Ida}, S. 1997, \apj, 477, 781

\bibitem[{Lovis} \& {Mayor} (2007)]{lovis}
{Lovis}, C., \& {Mayor}, M. 2007, \aap, 472, 657

\bibitem[{Marzari} \& {Weidenschilling} (2002)]{marzari}
{Marzari}, F., \& {Weidenschilling}, S.~J. 2002, \icarus, 156, 570

\bibitem[{Mayor} {et~al.} (2011)]{mayor}
{Mayor},  M., {Marmier},  M., {Lovis},  C., {et~al.}  2011, ArXiv e-prints,
arXiv:1109.2497

\bibitem[{Matsumura} {et~al.} (2013) ]{matsumura}
{Matsumura}, S., {Ida}, S., \& {Nagasawa}, M. 2013, \apj, 767, 129

\bibitem[{Morbidelli} (2005)]{morbidelli}
{Morbidelli}, A. 2005, ArXiv e-prints, arXiv:astro-ph/0512256


\bibitem[{Mr{\'o}z} {et~al.} (2017)]{mroz}
{Mr{\'o}z}, P.,{Udalski}, A.,{Skowron}, J., {et~al.} 2017, \nat, 548, 183

\bibitem[{Namouni} (2010)]{namouni}
{Namouni}, F. 2010, \apjl, 719, 145


\bibitem[{{Nesvorn{\'y}} {et~al.}(2007){Nesvorn{\'y}}, {Vokrouhlick{\'y}}, \&
  {Morbidelli}}]{nesvorny}
{Nesvorn{\'y}}, D., {Vokrouhlick{\'y}}, D., \& {Morbidelli}, A. 2007, \aj, 133,
  1962

\bibitem[{Nicholson} {et~al.} (2008)]{nicholson}
{Nicholson}, P.~D., {Cuk}, M., {Sheppard}, S.~S., {et~al.} 2008, The Solar System Beyond Neptune, University of Arizona Press


\bibitem[{{Payne} {et~al.}(2013){Payne}, {Deck}, {Holman}, \& {Perets}}]{payne}
{Payne}, M.~J., {Deck}, K.~M., {Holman}, M.~J., {et~al.} 2013, \apjl,
  775, L44

\bibitem[{{Peters} \& {Turner}(2013)}]{peters}
{Peters}, M.~A., \& {Turner}, E.~L. 2013, \apj, 769, 98

\bibitem[{Quinn} {et~al.} (1991)]{quinn91}  
{Quinn}, T.~R., {Tremaine}, S., {Duncan}, M. 1991, \aj, 101, 2287

\bibitem[{Rasio} \& {Ford} (1996)]{rasio}
{Rasio}, F.~A., \& {Ford}, E.~B. 1996, Science, 274, 954

\bibitem[{Rowan} {et~al.} (2016)]{rowan16} 
{Rowan}, D., {Meschiari}, S., {Laughlin}, G., {et~al.} 2016, \apj, 817, 104 


\bibitem[{Raymond} {et~al.} (2009)]{raymond09}
{Raymond}, S.~N., {Armitage}, P.~J., \& {Gorelick}, N. 2009, \apjl, 699, L88


\bibitem[{Raymond} {et~al.} (2010)]{raymond}
{Raymond}, S.~N., {Armitage}, P., \& {Gorelick}, N. 2010, \apj, 711, 772


\bibitem[{Raymond} {et~al.} (2011)]{raymond11}
{Raymond}, S.~N., {Armitage}, P., {Moro-Mart{\'i}n} A., {et~al.} 2011, \aap,
 530, A62

\bibitem[{Raymond} {et~al.} (2012) ]{raymond12}
{Raymond}, S.~N., {Armitage}, P.~J., {Moro-Mart{\'{\i}}n} A., {et~al.} 2012, \aap, 541, A11

\bibitem[{Raymond} \& {Izidoro} (2017)]{raymond17}
{Raymond}, S.~N. \& {Izidoro}, A. 2017, \icarus, 297, 134


\bibitem[{{Reynolds} {et~al.}(1987){Reynolds}, {McKay}, \&
  {Kasting}}]{reynolds}
{Reynolds}, R.~T., {McKay}, C.~P., \& {Kasting}, J.~F. 1987, Advances in Space
  Research, 7, 125

\bibitem[{Sasaki} {et~al.} (2012)]{sasaki}
{Sasaki}, T., {Barnes}, J.~W., \& {O'Brien}, D.~P. 2012, \apj, 754, 51

\bibitem[{{Sartoretti} \& {Schneider}(1999)}]{sartoretti}
{Sartoretti}, P., \& {Schneider}, J. 1999, \aaps, 134, 553

\bibitem[{{Scharf}(2006)}]{scharf}
{Scharf}, C.~A. 2006, \apj, 648, 1196

\bibitem[{{Simon} {et~al.}(2007){Simon}, {Szatm{\'a}ry}, \&
  {Szab{\'o}}}]{simon}
{Simon}, A., {Szatm{\'a}ry}, K., \& {Szab{\'o}}, G.~M. 2007, \aap, 470, 727

\bibitem[{Spalding} {et~al.} (2016)]{spalding}
{Spalding}, C., {Batygin}, K., \& {Adams}, F.~C. 2016, \apj, 817, 18

\bibitem[{Spergel} {et~al.} (2015)]{spergel}
{Spergel}, D., {Gehrels}, N., {Baltay}, C., {et~al.} 2015, ArXiv e-prints,
arXiv:1503.03757

\bibitem[{Sumi} {et~al.} (2011)]{sumi}
{Sumi}, T., {Kamiya}, K., {Bennett}, D.~P., {et~al.} 2011, \nat, 473, 349

\bibitem[{Teachey} {et~al.} (2017)]{teachey}
{Teachey}, A., {Kipping}, D.~M., \& {Schmitt}, A.~R. 2017, ArXiv e-prints, arXiv:1707.08563

\bibitem[{Tinney} {et~al.} (2011)]{tinney}
{Tinney}, C.~G., {Wittenmyer}, R.~A., {Butler}, R.~P., {et~al.} 2011, \apj, 732, 31

\bibitem[{Tremaine} (1993)]{tremaine93} 
Tremaine, S. 1993, Planets Around Pulsars, 36, 335 

\bibitem[{{Tremaine} {et~al.}(2009){Tremaine}, {Touma}, \&
  {Namouni}}]{tremaine09}
{Tremaine}, S., {Touma}, J., \& {Namouni}, F. 2009, \aj, 137, 3706

\bibitem[{{Veras} \& {Armitage}(2005)}]{veras05}
{Veras}, D., \& {Armitage}, P.~J. 2005, \apj, 620, L111

\bibitem[{{Veras} \& {Raymond}(2012)}]{veras12}
{Veras}, D., \& {Raymond}, S.~N. 2012, \mnras, 421, L117

\bibitem[{Ward} (1973)]{ward73}
{Ward}, W. 1973, Science, 181, 260

\bibitem[{Ward} (1974)]{ward74}
{Ward}, W. 1974, \jgr, 79, 3375

\bibitem[{{Williams} {et~al.}(1997){Williams}, {Kasting}, \& {Wade}}]{williams}
{Williams}, D.~M., {Kasting}, J.~F., \& {Wade}, R.~A. 1997, \nat, 385, 234

\bibitem[{Weidenschilling} \& {Marzari} (1996)]{weidenschilling}
{Weidenschilling}, S.~J., \& {Marzari}, F. 1996, \nat, 384, 619

\bibitem[{Winn} \& {Frabrycky} (2015)]{winn}
{Winn} N.~J., \& {Frabrycky} C.~D. 2015, \araa, 53, 409

\bibitem[{Wittenmyer} {et~al.}(2016)]{wittenmyer}
{Wittenmyer}, R.~A., {Butler}, R.~P., {Tinney}, C.~G., {et~al.} 2016, \apj, 819, 28


\end{thebibliography}
\end{document}